\documentclass[a4paper,fleqn,usenatbib]{mnras}

\usepackage{newtxtext,newtxmath}
\usepackage[T1]{fontenc}
\usepackage{ae,aecompl}

\usepackage{graphicx}	% Including figure files
\usepackage{amsmath}	% Advanced maths commands
\usepackage{amssymb}	% Extra maths symbols
\title[CDM subhalos in simulated ALMA observations]{Probing cold dark matter subhalos with simulated ALMA observations of macrolensed sub--mm galaxies}

\author[S. Asadi et al.]{
Saghar Asadi,$^{1}$\thanks{E-mail: saghar.asadi@astro.su.se (SA)}
Erik Zackrisson,$^{2}$
Emily Freeland$^{1}$
\\
$^{1}$Department of Astronomy, Stockholm University, Oscar Klein Center, AlbaNova, Stockholm SE--106 91, Sweden\\
$^{2}$Department of Physics and Astronomy, Uppsala University, Box 515, SE--751 20 Uppsala, Sweden
}

\date{Accepted XXX. Received YYY; in original form ZZZ}

\pubyear{2015}

\begin{document}
\label{firstpage}
\pagerange{\pageref{firstpage}--\pageref{lastpage}}
\maketitle

%% =========================== %%
%%          ABSTRACT           %%
%% =========================== %%
\begin{abstract}
  If the dark matter halos of galaxies contain large numbers of subhalos as predicted by the $\Lambda$CDM model, these subhalos are expected to appear in strong galaxy--galaxy lens systems as small--scale perturbations in individual images. We simulate observations of multiply--lensed sub--mm galaxies at $z\sim2$ as a probe of the dark matter halo of a lens galaxy at $z\sim0.5$. We present detection limits for dark substructures based on a visibility plane analysis of simulated Atacama Large Millimeter/submillimeter Array (ALMA) data in bands 7, 8 and 9. We explore two effects: local surface brightness anomalies on angular scales similar to the Einstein radius and the astrometric shift of  macroimages.  This improves the sensitivity of our lens modeling to the mass of the lens perturber. We investigate the sensitivity of the detection of low--mass subhalos to the projected position of the subhalo on the image plane as well as the source structure and inner density profile of the lens. We demonstrate that, using the most extended ALMA configuration, pseudo-Jaffe subhalos can be detected with 99\% confidence down to $M = 10^7 M_\odot$. We show how the detection threshold for the three ALMA bands depends on the projected position of the subhalo with respect to the lensed images and conclude that, despite the highest nominal angular resolution, band 9 provides the poorest sensitivity due to observational noise. All simulations use the \emph{ALMA Full ops most extended} ALMA configuration setup in \textsc{casa}.
\end{abstract}

% Select between one and six entries from the list of approved keywords.
% Don't make up new ones.
\begin{keywords}
techniques: interferometric -- gravitational lensing: strong -- galaxies: dwarf -- radio continuum: galaxies -- submillimetre: galaxies -- dark matter
\end{keywords}

%%%%%%%%%%%%%%%%%%%%%%%%%%%%%%%%%%%%%%%%%%%%%%%%%%

%%%%%%%%%%%%%%%%% BODY OF PAPER %%%%%%%%%%%%%%%%%%

%% =========================== %%
%%          INTRO              %%
%% =========================== %%
\section{Introduction}
\label{sec:introduction}
Cosmological N--body simulations predict the existence of dark matter halos spanning the mass range from galaxy cluster halos ($\sim 10^{15} M_\odot$) down to the smallest halos corresponding to the cutoff in the primordial matter power spectrum. {The high--mass end of the halo mass function can be tested through anisotropies in the cosmic microwave background radiation \citep[e.g.][]{COBE+1992, WMAP9, Planck2014}.  The low--mass end is highly dependent on the detailed properties of dark matter particles, particularly particle mass and decoupling velocity \citep[e.g.][]{Bertone+2005}.  The cutoff mass constraint varies widely from one particle candidate to another. For example, Weakly interacting massive particles (WIMP) models usually suggest a minimum halo mass as low as $\sim 10^{-11} M_\odot$ for light dark matter particles \citep[e.g.][]{Bringmann2009, Cornell+2013}}.

Dark matter--only cosmological N--body simulations assume the gravitational interaction between WIMPs to be the dominant driving force behind structure formation \citep[see, for instance,][for projects Aquarius, Phoenix, and EAGLE, respectively]{Springel+2008, Gao+2012, Schaller+2015}. { As a result, dark matter halos form in a hierarchical process during which small halos that are bound to the gravitational potential of a massive halo join the smooth matter content of the parent halo through tidal disruption. This gradual process takes up to several billion years to complete. During this hierarchical assembly subhalos remain within the parent halo in the form of a clumpy substructure on top of the smooth matter distribution}. A generic prediction of dark matter--only simulations is that subhalos represent $\sim10\%$ of the total halo mass of a galaxy at $z = 0$ \citep{Gao+2008, Maciejewski+2011}. These subhalos of simulated galaxy--sized CDM halos are best--fit by a mass function of the type:

\begin{equation}
\frac{\mathrm{d}N}{\mathrm{d}M}\propto M^{-\alpha} \nonumber
\end{equation}
with $\alpha\approx 1.9$ \citep[][]{Springel+2008, Gao+2008, Gao+2012, Xu+2012} at the low--mass end.

Comparisons of the dark matter halo mass function with observations of galaxies find that the numbers of simulated dark halos greatly exceed the number of observed dwarf satellites of Milky Way and Andromeda \citep[see e.g.][]{Klypin+1999,Moore+1999}.  There is also a mismatch between the most massive subhalos at $z=0$ in simulations and the most massive satellite galaxies in the local group \citep{Boylan-kolchin+2011, Boylan-kolchin+2012}
%A relatively recent challenge for $\Lambda$CDM at small--scales is the mismatch between the most massive subhalos at $z=0$ in simulations and the most massive satellite galaxies in the local group, the so--called ``too big to fail'' problem \citep{Boylan-kolchin+2011, Boylan-kolchin+2012}.

It has been suggested that the halo of the Milky Way is less massive than previously estimated and that therefore, the observed local massive dwarfs should be compared to simulated subhalos of a less massive halo \citep{Boylan-kolchin+2012}. However, similar discrepancies have been reported within the local group \citep{Kirby+2014, Garrison-Kimmel+2014, Tollerud+2014} as well as in field galaxies \citep{Papastergis+2015}, suggesting that a more general solution is required. For example, \citet{Vera-Ciro+2013} show that the solution may lie with the mass profiles of the simulated subhalos associated with bright observed satellites of the host galaxy. Additionally, in the absence of full hydrodynamic cosmological simulations, adopting baryonic physics and feedback processes in simulations on galaxy to galaxy group scales addresses both the number of halos and the mismatch at the high mass end \citep[e.g.][]{Garrison-Kimmel+2014,Sawala+2014, Fry+2015, Schaller+2015, Despaliandvegetti2017}. Adopting non--standard properties for dark matter particles also provides a competitive and complimentary solution to both discrepancies \citep[e.g.][]{Lovell+2012, Vogelsberger+2012, Wang+2014}.

Simulated dark matter halos generally fit to a universal 3D density profile increasing towards the center as $\rho\propto r^{-\alpha}$, where $\alpha=1$ (shallower than an isothermal sphere with $\alpha=2$), while the logarithmic slope decreases at larger radii, i.e. $\alpha=3$ for $r>r_s$ \citep{NFW96}. Dynamical measurements of stellar and gas content in the central kpc of dwarf and low surface brightness galaxies favor a constant density core (i.e. $\alpha=0$) for these regions \citep[e.g.][]{Moore1994, Zackrisson+2006, Kuziodenaray+2008, Oh+2011, Kuziodenaray+2011, deBlok2010, Walkerandpenarrubia2011, Amoriscoandevans2012}. However, measurements of this type are sensitive to the method used to derive the dynamical mass
contribution from the dark matter, which requires precise modeling of stellar populations and the mass function in these galaxies \citep{BreddelsandHelmi2013, Strigari+2014}.

\begin{figure}
\includegraphics[width=1.0\columnwidth]{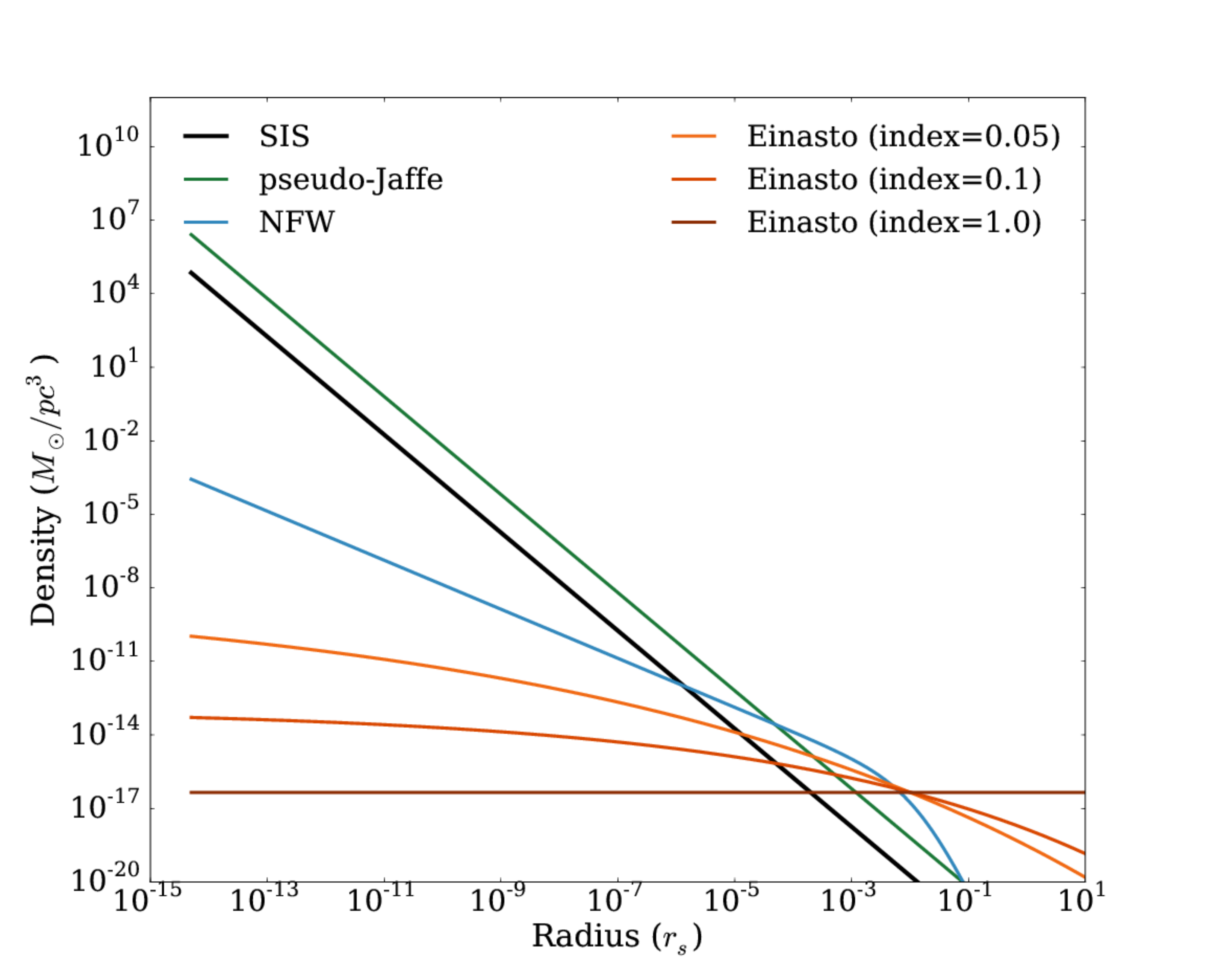}
\caption{Six different inner density profiles for dark matter halos; Singular isothermal sphere (SIS) shown in black, the pseudo-Jaffe profile in green, the standard CDM halo profile suggested by NFW in blue, and the three--parameter Einasto profile with different Einasto indices in shades of brown. All profiles are normalized to the virial subhalo mass $M_\mathrm{vir} = 10^8 M_\odot$ and the \citet{Bullock+2001} recipe is adopted to set the concentration parameter $c_\mathrm{vir}$ at $z = 0.5$. Investigating $\alpha_\mathrm{Ein}$ in a range between $10^{-2} - 10^{-1}$ is relevant in the context of gravitational lensing. However, the Einasto index depends systematically on mass, meaning that the term ``universal'' density profile needs to be used keeping that in mind that density profiles of halos with various masses are not self--similar.}
\label{fig:profiles}
\end{figure}

Baryonic feedback processes such as adiabatic contraction, gas outflows, photoionization from the ultraviolet background \citep{Sawala+2014} and even dynamical cosmic ray feedback \citet{Chen+2016} have been suggested to resolve the ``cored'' central density profiles in baryon--dominated regions \citep[see e.g. ][]{Navarro+1996, deBlok+2001, Maxwell+2015}.  However, observations of dark matter--dominated halos seem to be following a nearly universal density profile which is much ``cuspier'' in the center (see Figure \ref{fig:profiles})

Gravitational lensing can be a powerful probe of dark halo substructure, corresponding to the low--mass end of the dark matter halo mass function at $z>0$ \citep[for a review see \citet{Zackrissonandriehm2009, Vegetti+2010, Li+2016} and for some attempts at doing so see e.g.][]{Vegetti+2012, Vegetti+2014, VegettiandVog2014, Hezaveh+2016}. 

The strong lensing effect requires a close alignment of the source, the lens, and the observer to result in multiple lensed images of the background source. The total mass content of the foreground lens, including both the underlying smooth dark matter halo and the substructure, is probed by studying locations and relative fluxes of these macroimages. Observations have attempted to constrain the contribution from dark matter substructure in compound lensing systems using two techniques: the statistical examination of flux ratio anomalies which is more sensitive to a population of low/medium mass subhalos, and surface brightness anomalies made by individual subhalos. 
While the latter effect is not sensitive to low--mass substructure (given current data quality), it does have the merit that the surface brightness anomalies produced by individual substructure are likely to be distinguishable even in a case where the main lens contains more than one massive substructure.

In a strong lens system, positions of lensed images and their flux ratios with respect to each other are tightly bound to the lens solution for the main lens in the system. Therefore, any deviation in the measured flux of lensed images of the quasar are attributed to additional effects, including the presence of dark substructures in the main lens \citep{Chiba2002, Keeton2003, Kochanekanddalal2004, Nierenberg+2014, Cyr-Racine+15} or along the line of sight \citep[see e.g. ][]{McCully+2016}, or even propagation effects \citep{Xu+2015, Hsueh+2016}. A more detailed discussion of this can be found in section \ref{subsec:low_redshift_line_of_sight_contamination_and_external_shear_vs_central_density_profile}.

If flux ratio anomalies are indeed interpreted as evidence of substantial small--scale structure within the main lens, they provide statistical constraints on the mass fraction of galactic halos in the form of substructure and possibly the mass function of the subhalos. The mean projected subhalo mass fraction based on N--body simulations is estimated to be $f_\mathrm{sub}\approx 0.005$, although with a scatter of a factor of $\sim 10$ among different halos in the Aquarius simulation \citep{Xu+2009}.  The same fraction estimated from observations of individual millilensing effects is a mean fraction of % $f_\mathrm{sub}\approx 0.007$ -- the reported mean fraction is
 $f_\mathrm{sub}\approx 0.0076^{+0.0208}_{-0.0052}$ at 68\% confidence level \citep{Vegetti+2014}. More recently, \citet{Hezaveh+2016} use ALMA observations of SDP81 to estimate the projected subhalo mass fraction within 10 kpc of the lens galaxy as 0.003 to 0.0035 at 90\% confidence. It is crucial to note that while \citet[][]{Xu+2009, Xu+2010} consider only Milky Way sized halos from the Aquarius project, observations constrain only massive early--type galaxies. The subhalo abundance is expected to increase rapidly with the host halo mass. Therefore, the strong dependence of the projected subhalo fraction on the redshift, mass, and ellipticity of the host halo cannot be neglected \citep[see e.g.][]{MetcalfandAmara2012, Xu+2015}. Moreover, CDM substructure is unlikely to be the only cause of these flux ratio anomalies. This implies that the estimated subhalo fractions based on this assumption should only be considered as an upper limit \citep[see e.g.][who are investigating two classes of such contributions; interstellar propagation effects and the use of improper models for the main lens that are either too simple or unrealistic]{Xu+2015}. 

In this paper we seek to derive mass detection limits for subhalos as a function of their projected location in the image plane observations using ALMA bands 7, 8, and 9. We also investigate the effect that the choice of the halo density profile has on recovering the mass and position of the subhalo in our simulations.

In the next section, we discuss the two different groups of empirical halo density profiles. Section \ref{sec:method} elaborates on the details of our simulations, including the choices of sources and perturbers and the use of complex visibilities and a comparison to image plane modeling. Our results are presented in section \ref{sec:results} and finally section \ref{sec:discussion} discusses the results of this work in the context of the field as well as the limitations. A summary of the results can be found in section \ref{sec:summary}

\section{Different forms of halo substructure}
\label{sec:different_forms_of_halo_substructure}
The central slope of the density profiles of dark matter halos can be measured both from observational data and fits to halos in N--body simulations. In this regard, the single--parameter (cored) singular isothermal sphere (ellipsoid) profile provides an acceptable lens model for the mean dark matter halo of galaxies. On the other hand, the universal density profiles of field halos in CDM simulations can be reasonably well described by \citet[][ hereafter NFW]{NFW96} profile:
\begin{equation}
\label{eq:NFW}
\rho(r)=\frac{\rho_\mathrm{s}}{(r/r_s)(1+r/r_s)^{2}} \nonumber
\end{equation}
where $r_s$ is the characteristic scale radius of the halo, i.e. the radius at which $\rho \propto 1/r^2$, and $\rho_s$ is the density at $r=r_s$. An extra parameter called the \emph{concentration parameter} $c$ relates the scale density of the halo to its virial radius and is defined as $c\equiv r_s/r_\mathrm{vir}$. The concentration parameter therefore, contains information about the formation and evolution of the halo and depends on the time of the collapse of the halo as well as its virial mass. Given the hierarchical formation of halos, the low--mass halos were formed at higher redshift where the mean density of the Universe was higher and so was the inner density of collapsed halos. This results in a weak $c-M_\mathrm{vir}$ correlation such that the concentration parameter decreases with increasing $M_\mathrm{vir}$. {Additionally, low--mass subhalos gradually lose mass in tidal interaction with the parent halo which leads to a further increase of their $c_\mathrm{vir}$ with time \citep{Bullock+2001, Maccio2008}.}

Relaxing the central logarithmic slope $\gamma = \frac{d\ln(\rho/\rho_s)}{d\ln(r/r_s)}$ in the basic two--parameter form of NFW profile provides a better fit to individual halos in CDM simulations. In this three--parameter form, the inner cusp slope becomes progressively shallower towards the center, eventually reaching inner slope of $\gamma\geq-1$. The generalized NFW profile (gNFW) is formulated as:

\begin{equation}
\rho(r) = \frac{\rho_s}{(r / r_s)^\gamma(1 + r/r_s)^{3-\gamma}} \nonumber
\end{equation}
where $\gamma = 1$ gives the traditional NFW profile and $\gamma = 2$ is equivalent to a singular isothermal sphere (SIS). 

Another option for a three--parameter profile is the Einasto profile, inspired by the two--dimensional Sersic surface brightness profile of elliptical galaxies. Both high-resolution measurements of central stellar and gas content of low surface brightness dwarf galaxies, and high--resolution CDM simulations, tend to indicate more consistency with a three--parameter density profile rather than the traditional NFW ones. There are various studies suggesting that simulated CDM halos are better described by a three--parameter model such as the Einasto profile than the standard NFW \citep[e.g.][]{Navarro+2004, Gao+2008, DiCintio+2014, Duttonandmaccio2014, Katz+2016}. The extra parameter describing dark halo density profiles, $\alpha_\mathrm{Ein}$, gives the density profile more flexibility in shape, i.e. $\gamma(r) = -dln\rho/dlnr$. The three--dimensional Einasto profile takes the form:
\begin{equation}
\ln\left(\frac{\rho(r)}{\rho_s}\right) = -b\left[\left(\frac{r}{r_s}\right)^{\frac{1}{n}} - 1\right] \nonumber
\end{equation}
where the \emph{Einasto index} $\alpha_\mathrm{Ein} = \frac{1}{n}$, and therefore the logarithmic slope becomes $\gamma = -\frac{b}{n}(\frac{r}{r_s})^\frac{1}{n}$. 

Best--fit density profiles to simulated dark matter halos of a variety of masses in the Aquarius project show inner slopes shallower than the original NFW \citep{Navarro+2010}. These halos are not self--similar, i.e. Einasto index changes with halo mass. \citet{Navarro+2004} find the Einasto index for halos in the mass range between dwarves and clusters to be 0.12--0.22, with an average value of 0.17. According to \citet{Hayashiandwhite2008, Gao+2008}, $\alpha_\mathrm{Ein}$ tends to increase with mass and redshift in halos of the Millennium simulation. From the gravitational lensing point of view, Einasto profiles are more demanding to work with as one cannot derive an analytical surface mass density as a function of $\alpha_\mathrm{Ein}$. Hence the lens equation needs to be solved numerically for each case.

A pseudo-Jaffe density profile is one of the most popular models for dark matter halos. While the original ellipsoidal Jaffe profile reads as $\rho \propto r^{-2} (r + r_\mathrm{trun})^{-2}$ \citep{Jaffe1983} resulting in a rotation curve declining for $r<r_\mathrm{trun}$, the more commonly--used pseudo--Jaffe profile used for lensing purposes has a radial profile of the form $\rho \propto r^{-2} (r^2 + r_\mathrm{core}^2)^{-1}(r^2 + r_\mathrm{trun}^2)^{-1}$ which is more convenient for analytical lensing calculations and leads to a smoother break at $r_\mathrm {trun}$ \citep{Keeton2001}. The surface density for this model will then have the form:
\begin{equation}
\Sigma = \kappa [(r^2 + r_s^2)^{-1/2}(r^2 + r_\mathrm{trun}^2)^{-1/2}] \nonumber
\end{equation}
This model also matches the standard singular isothermal ellipsoid (SIE) model for $r\ll r_\mathrm{trun}$ with a finite total mass.

%% =========================== %%
%%          METHOD             %%
%% =========================== %%

\section{Method}
\label{sec:method}

The Einstein radii of lens perturbers with $\rho(r) \propto r^{-2}$ (SIS or pseudo--Jaffe profiles) and virial masses corresponding to a dark matter substructure mass range $M_\mathrm{vir} < 10^{10} M_\odot$ are over an order of magnitude too small to be directly detectable even with milliarcsecond resolution \citep[see Figure 1 in][]{Riehm+2008}. However, as a result of the shear made by the lens perturber, the corresponding surface brightness perturbations appear at a range of angular scales larger than $R_E$.

Here we present lensing simulations which investigate parameter estimation in the case of strong lens systems with a secondary lens perturber as observed with ALMA bands 7 to 9. The parameters we seek to recover are the mass and projected position of the lens perturber. We also investigate how assuming different halo density profiles influences the substructure mass estimate. The lensing simulations employ Glafic, a grid--based software which solves the gravitational lensing equation and which offers several lens and source models \citep{OguriGlafic}. 

\subsection{Strongly--lensed sources}
\label{subsec:strongly_lensed_sources}
There are a few criteria to fulfill when searching for strongly--lensed targets suitable for millilensing by halo substructures. One is the achievable angular resolution and the other the covered area in the lens plane probing the potential substructures of the lens. Optimizing the combination of the two leads to different combinations of source and observing frequency which are limited by instrumental capabilities. 

Previous work by \citet{Zackrisson+2013} investigates the prospects of halo substructure detection using synchrotron emission by blazars observable at high angular resolutions by various VLBI arrays at frequencies between 8.4 -- 86 GHz. The synchrotron radiation from blazars is emitted within compact regions much smaller than a parsec, extending to kpc scales only at very low frequencies, while a star--forming galaxy emits thermal dust continuum across a physical region a few orders of magnitude more extended. As concluded in \citet{Zackrisson+2013}, in addition to very high angular resolution observations for both standard CDM halos and compact alternative substructures, a larger coverage of the source plane is needed to boost the probability of suitable lens--source alignment to sample even the massive end of the subhalo mass function. 

This work employs simulations of continuum emission from multiply--lensed sub--mm galaxies (SMGs) to investigate the prospects of detecting dark halo substructures. Sub--millimeter galaxies have a redshift distribution which peaks at $z\simeq2$ \citep[][]{Wardlow+2013} and typical flux densities of $S_{850} \geq 5$ mJy \citep[][]{Karim+2013}. The study of lensed SMGs with angular resolutions on the order of $\sim 10$ mas can provide spatial resolutions of a few 100 pc in the source plane at $z\sim2$, which in turn corresponds to a spatial resolution of a few $\times$ 10 pc in the lens plane at $z \sim 0.5$. This is sufficient to probe dwarf galaxies and dark substructure in the galactic halo. The percentage of SMGs expected to be strongly--lensed with an average magnification factor of $\mu \sim 9$ is 32--74\% (for $S_{500\mu\mathrm{m}} \geq 100$ mJy) and 15--40\% (for $S_{500\mu\mathrm{m}}$ between 80-100 mJy) \citep[][]{Wardlow+2013}.

\citet{Blain+1999} compare the unlensed SMG number counts at 450 and 850 $\mu$m -- roughly corresponding to ALMA band 9 and band 7, respectively. Although the data are limited, the population model suggests more numerous sources of the same flux density at 850 $\mu$m, while those detected at 450$\mu$m are fewer but more luminous. In the context of strong galaxy--galaxy lensing, this implies better chances of suitable strong lensing alignments at lower frequency bands. This, and the results presented in section \ref{sec:results}, makes ALMA band 7 and 8 observations preferred over those of band 9 for the purpose of this work.

\subsection{Lensing simulations}
\label{subsec:lensing_simulations}
\begin{figure*}
\includegraphics[width=0.9\textwidth]{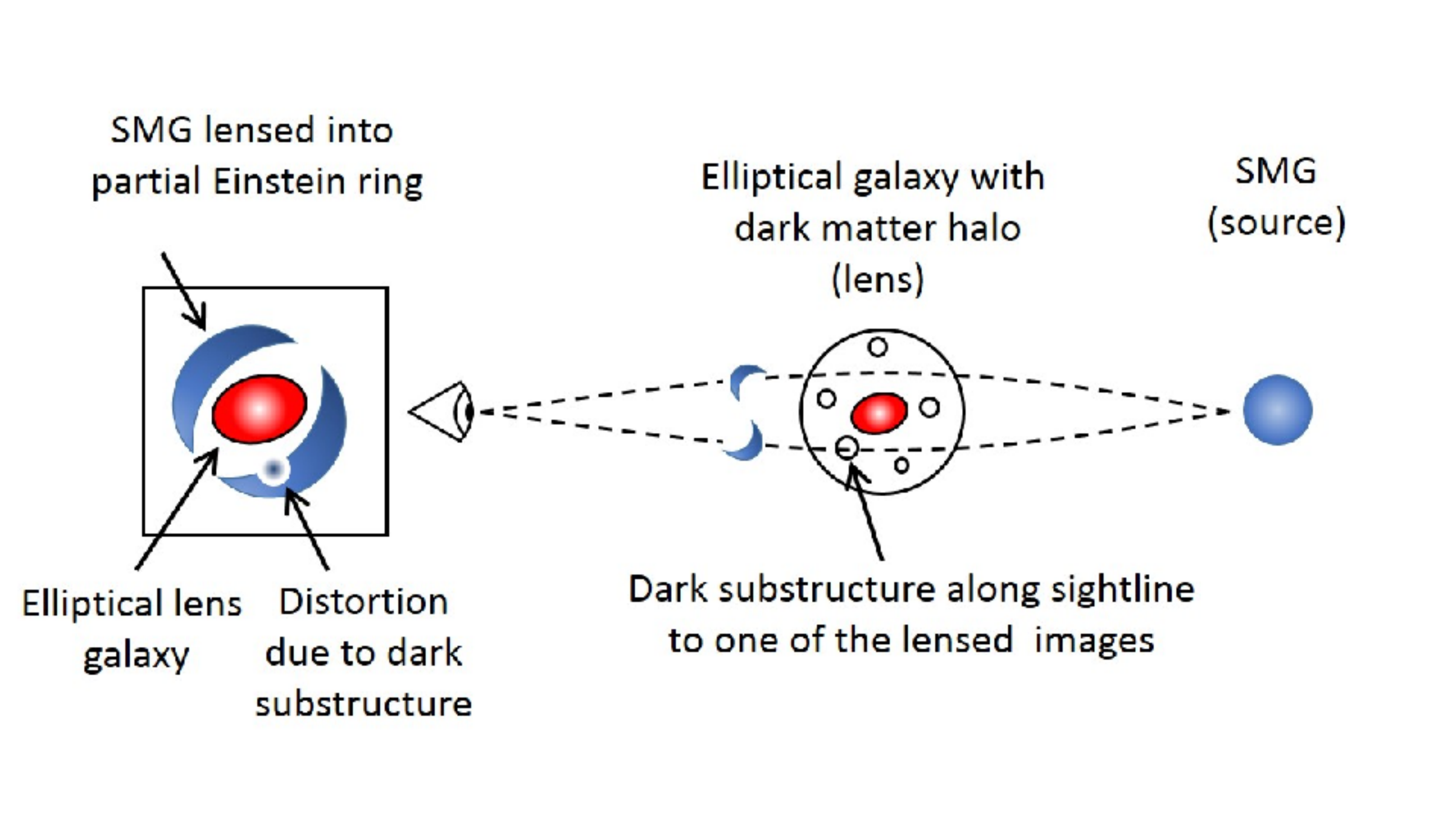}
\caption{A schematic view of the spatial setup of the lens system. The source (sub--mm galaxy at $z\simeq2$) is multiply--imaged by the main lens. If the dark matter halo of the lens galaxy is populated with substructures, a subhalo along the line of sight towards one of the macroimages can cause a secondary lensing effect. Given a strong--enough secondary lensing due to the subhalo, this distortion will affect only one of the macroimages, leaving the others intact, unlike SMG source structure that replicates in all macroimages.}
\label{fig:lensing_setup}
\end{figure*}

To simulate the effects of dark halo substructure on macrolensed SMGs we use a numerical scheme similar to that developed by \citet[][]{Metcalfandmadau2001}. A schematic view of the lens system is illustrated in Figure \ref{fig:lensing_setup}. An extended source is assumed to be multiply--imaged by a foreground galaxy. The lens equation is used to determine the lens plane positions of the corresponding macroimages. A secondary lens perturber (substructure) is then placed at different positions on the lens plane. The deflection angles (with contributions both from the subhalo and the macrolens) are computed for every pixel within this region and converted into a numerical surface brightness map of the macroimage. These maps are initially generated with a very fine pixel scale (0.005 arcsec) and are then run through \textsc{casa} to create the expected visibilities achieved with ALMA in its most extended configuration. This extended configuration has a maximum baseline $\simeq$ 16 km and provides an angular resolution between $7 - 10$ mas depending on the observational frequency band (band 7: $275-370$ GHz, band 8: $385-500$ GHz, and band 9: $602-720$ GHz). 

\begin{figure}
\includegraphics[width=0.5\textwidth{}]{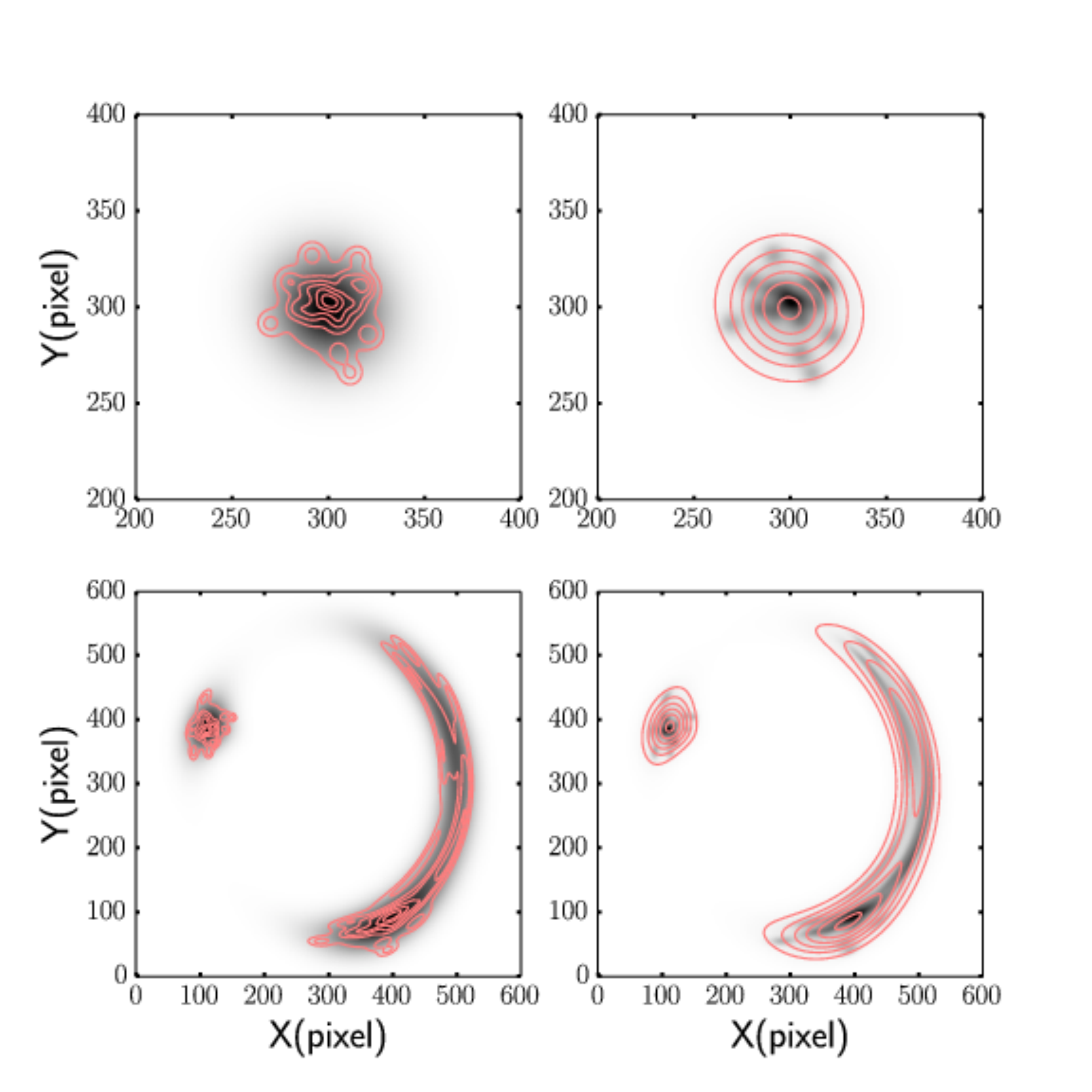}
\caption{The two extreme cases of smooth vs clumpy source model (with structures on the scale of the image separation made by secondary lenses), and the resulting lensed images of the source in a \emph{double} configuration. On the left column, the smooth source, and the corresponding lensed image are mapped in gray scale, and the red contours show the emission from the clumpy source. On the right column, the clumpy source is in gray scale and the smooth source is shown by the overlaid contours.}
\label{fig:sources}
\end{figure}

Physical sizes of SMGs are between 4--8 kpc \citep{Swinbank+2008}. In order to explore the effect of source structure on our detections we compare two source models as shown in Figure \ref{fig:sources}. In one case, referred to as the smooth source model, the dust continuum emission is modeled with a single Gaussian component with FWHM $\simeq 0.5-1.0$ arcsec and an unlensed flux density of 5 mJy at $z=2.0$. In the other case, the clumpy source model, a source of the same size and integrated flux density is assumed to be composed of identical randomly positioned components. These clumps are intended to represent star forming regions. These regions are estimated to be $\le $200 pc across \citep{Swinbank+2010}. This corresponds to $\le $0.02 arcsec at z$\simeq$ 2. Given the relatively strong continuum emission from these sources, the required integration time with the full array does not exceed 2 hours, regardless of the details of the source structure. The free parameters related to the source in the fitting process are the position, flux density, and size (for the smooth source) or number of clumps (in case of the clumpy source).

\begin{figure*}
\includegraphics[width=1.0\textwidth{}]{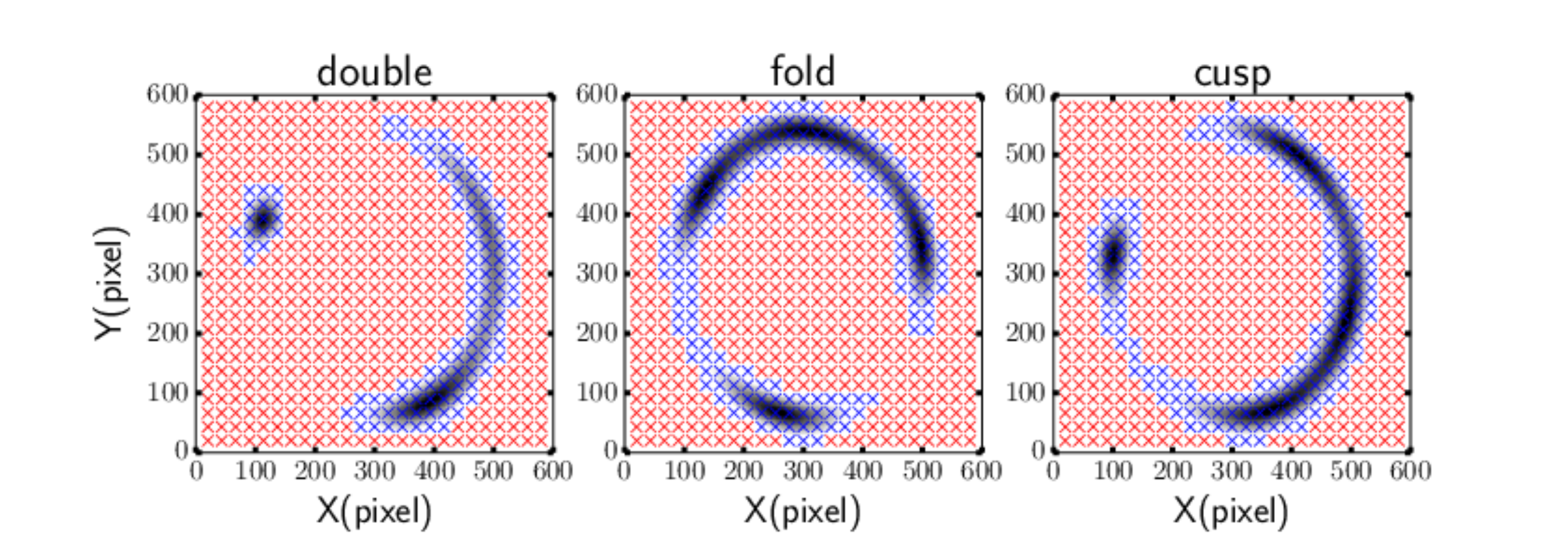}
\caption{Three instances of different lens system configurations; double, fold and cusp from left to right. Each simulated sky region is divided into grid locations (25 $\times$ 25), the centers of which are shown by red crosses. The blue crosses correspond to grids that are brighter than 3\% of the maximum surface brightness of each source assuming that the detectable parts of the lensing system are down to $\sim 10\%$ of the peak. The blue crosses also correspond to the locations at which substructures are inserted when testing how their detectability depends on position. The test substructure positions remain unchanged in the case of simulations with a clumpy source, for comparison.}
\label{fig:sim_preparation}
\end{figure*}

The simulations presented in this paper are all based on lens--source configurations that in the absence of substructure would give typical magnifications $\mu_1\approx 10$ for at least one of the macroimages. This is consistent with typical magnification for observed SMGs \citep[e.g.][]{Wardlow+2013}. For each realization, the main lens is modeled as a generalized SIE with $R_\mathrm{Ein} \sim 1$ arcsec and at least one macroimage with a magnification $\approx 10$ for a full range of ellipticities and position angles (free lens parameters in the fitting process) at redshift 0.5. This implies that the central region of the main lens is assumed to be dominated by the baryonic component of the host halo which is consistent with lensing observations \citep[e.g. ][]{Koopmans+2006, Sluse+2012} and other simulations of strong lens systems by early--type galaxies \citep[e.g. ][]{Hezaveh+2013b, Xu+2015}. We add external shear to account for possible perturbations along the line of sight due to objects at redshifts equal or smaller than the main lens. In accounting for the external shear, we follow what is assumed in \citep{Keeton2003} with a median $|\gamma| = 0.05$ and allow for a full range of orientations. Our simulations consist of systems with three different types of lens configurations - cusp, folds, and doubles - and consider 5 different realizations of each configuration. In each case, the simulated region of the image plane ($\sim 9$ arcsec$^2$) is divided into a grid of 25 by 25 pixels. Those grids whose center would correspond to a sky brightness of $\geq 3\%$ of the maximum of the model are chosen to populate with substructures. Only a single substructure per simulation (see Figure \ref{fig:sim_preparation}) is assumed. This results in over 100 different secondary lens positions for each lens system.

Each of these systems -- consisting of one primary and one secondary lens -- are first simulated by Glafic then fed to \textsc{casa}. The resulting visibilities are then weighted and each weighted visibility set is used to estimate the following parameters of the lenses while the source parameters are fixed; mass, position, ellipticity, and position angle for the primary lens, external shear in the system, $|\gamma|$ and $\theta_\gamma$, mass, and position (x and y) of the secondary lens.  The secondary lens is assumed to have a pseudo--Jaffe profile. 

The $\chi^2$ minimization is done in two steps, first the randomly initialized guess goes through a Nelder minimizer, the solution of which is fed to a Levenberg--Marquardt minimizer to estimate the posterior distribution of model parameters (including those of the lens(es) and the source). This process is performed for the combination of the two weighting schemes and each cycle is repeated at least 1000 times with random initial guesses. The parameter uncertainties are derived from the resulting probability distribution functions.

To confirm the detection limits, we cross validate the results of three means of measuring residual emission on the source plane. First, we examine the sky model from the original simulations to which \textsc{casa} adds additional telescope and noise effects.  Second, we examine the smoothed visibility function (phase and amplitude measurements from \textsc{casa} as a function of uv--distance). These visibility measurements are noisy and need to be smoothed before comparison. And third, we examine the CLEAN images of the simulated ALMA observations. We ensure that no system is marked as detectable in cases 2 and 3 (especially for clumpy source models) without the original simulation (without noise) being marked detectable. It is always the case that the systems that qualify as detectable in case 1, but not 2 or 3 are marked undetectable.

\subsection{Visibility modeling}
\label{subsec:visibility_modeling}
In radio interferometry, the directly measured sky values are complex visibilities located at various (u,v) distances. Each visibility point carries information about flux from all over the sky. As a result of the Fourier transformation, narrow features on the (u,v) plane transform into wide features on the image plane and vice--versa. Therefore, surface brightness maps (images) made from interferometry data are subject to various assumptions in the deconvolution process. The resulting images of the same visibility set can differ notably in featured structures. These differences depend on the prior knowledge of the user about the source and the chosen weighting scheme. Pixel--to--pixel errors on the image plane are correlated and affected by the choice of grid cells in the imaging process. Therefore, direct visibility modeling is favored over surface brightness map to model comparison \citet[see e.g. \citep{Wucknitz2004, Bussmann+2012, Hezaveh+2013b, Rybak+2015}][ for different visibility modeling methods]{}. In cases where the source can be reduced into simple analytical components, e.g. a point source, Gaussian profile, uniform disk, circular ring, etc., or a combination of them, the best fit visibility model can be found by surveying the parameter space for each component. However, if the flux density distribution of the source is more complicated, then an initial flux density distribution in the image plane is made, then the calculated visibilities of that initial map are compared to the observed values. This method requires calculating the visibilities of all investigated points in the parameter space, and fitting measured visibilities to those of the models, and therefore is more computationally involved than image plane modeling.

To account for lensing effects on both large and small angular scales, the commonly--used \emph{natural} and \emph{uniform} weighting schemes are used simultaneously to fit the visibilities. Even by changing the weighting scheme, in order to do a pixel--to--pixel comparison, the (u,v) plane cell size needs to remain unchanged. For fulfilling the criterion of Nyquist sampling, $\Delta x < \frac{1}{u_\mathrm{max}}$, the minimum (u,v) grid sizes for band 7, 8, and 9 are $\sim 0.006$, $\sim 0.005$, and $\sim 0.002$ respectively. In all cases, an integration time of $\leq 6$ s is adopted to avoid information loss in phases, given that the luminous ring is about an arcsecond away from the phase center.

Sampling of the (u,v) plane tends to be denser near the origin, giving a better signal to noise ratio for short--spacing measurements. Normal beam convolution corresponds to a sampling/weighting function in the (u,v) plane that usually depends on the radial distance from the origin on the (u,v,) plane. Therefore, giving all datapoints the same weight (as in the \emph{natural} weighting scheme), leaves the relative contribution of data to the sampling function. With the emphasis on short--spacings, the thermal rms noise level of the resulting map gets minimized, as the synthesized beam gets broader. This means that the small--scale structure of the source gets washed out by beam convolution and is an undesirable effect for a composite source. The alternatives, for preserving most of the structure in the source, at the cost of sensitivity, are to either set a minimum (u,v) limit, or downweight the data points close to the origin on the (u,v) plane. The other commonly--used weighting scheme, known as \emph{uniform} weighting is simply designed to weight visibilities at various (u,v) distances inversely proportional to their abundance. The weighting in this scheme depends on griding of the (u,v) plane, where each data point is weighted according to

\begin{equation}
w_k = \frac{1}{N_s(k)} \nonumber
\end{equation}
where $N_s(k)$ is the number of data points within a symmetric region (cell) around the $k^{th}$ data point on the (u,v) plane.
The visibility plane is usually gridded into regularly--spaced square cells to ease the use of fast fourier transform algorithms. However, since data points are not regularly--spaced in the visibility plane, an interpolation operation is required.

Simulated ALMA visibilities are made using \textsc{casa} software with the addition of atmospheric thermal noise by setting \verb+thermalnoise="tsys-atm"+ in \verb+simobserve+. The same weighting recipes as used in the \textsc{casa} \verb+clean+ task have been applied to simulated visibilities on the visibility plane. 

%% =========================== %%
%%          RESULTS            %%
%% =========================== %%
\section{Results}
\label{sec:results}

\begin{figure*}
\includegraphics[width=0.45\textwidth]{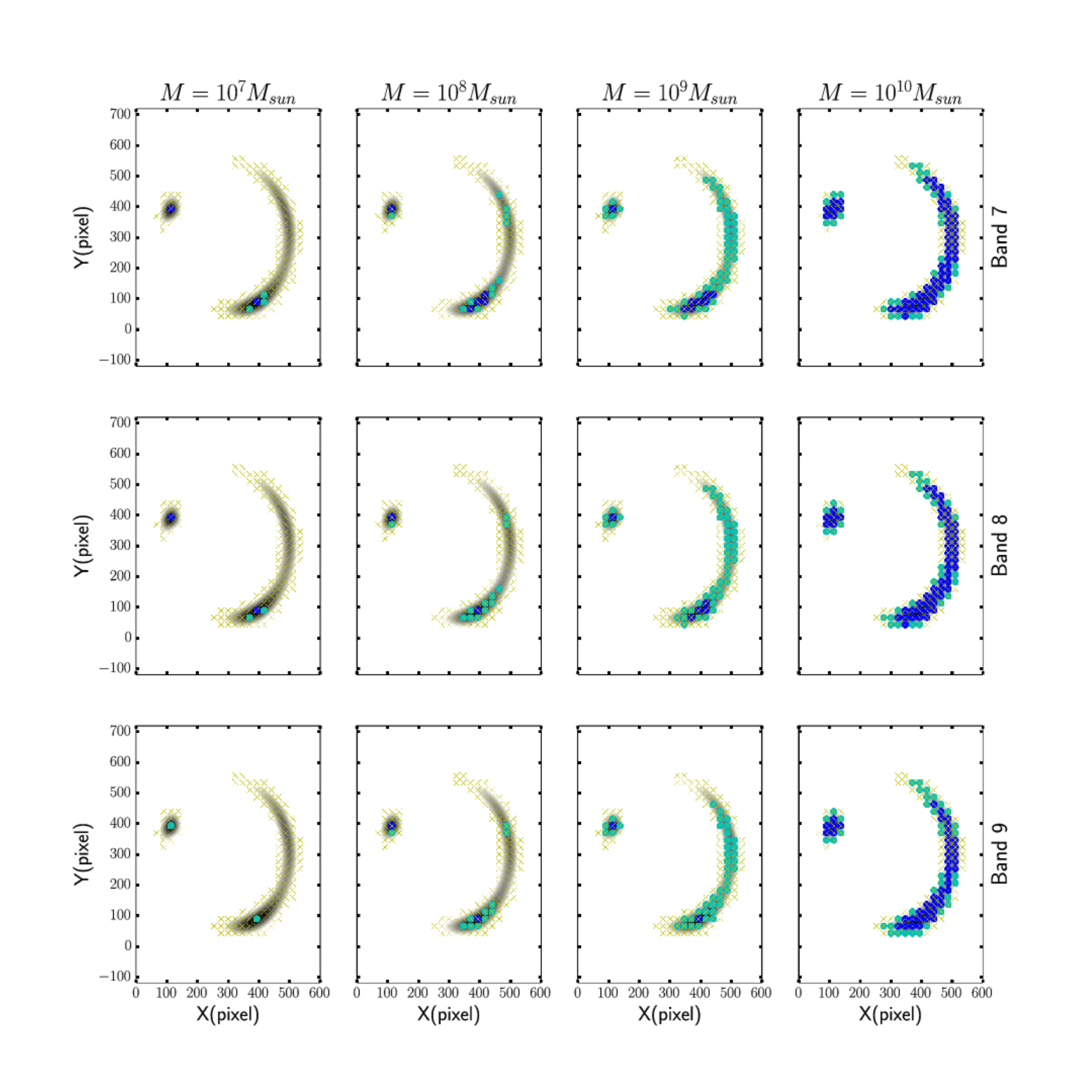}
\includegraphics[width=0.45\textwidth]{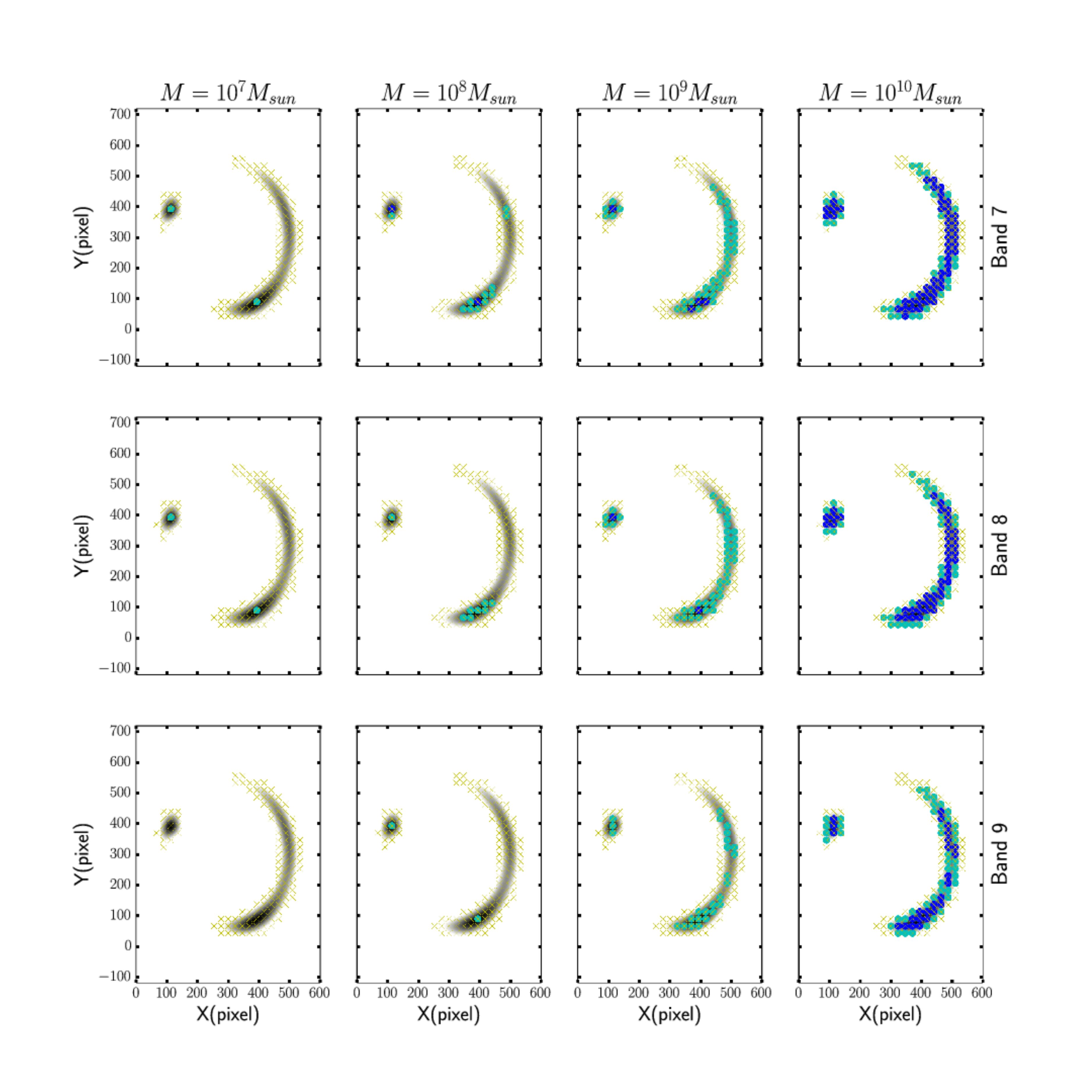}
\caption{Detectability expectation as a function of position of the substructure for SIE substructures of mass $10^7 - 10^{10} M_\odot$ (left to right) and a double lens system using ALMA bands 7, 8, and 9 (from top to bottom). The dark blue points are indicative of a recovered mass and position of the substructure at $10\sigma$ confidence. The cyan regions depict the subhalo positions at which the mass and position of the secondary lens is correctly recovered at $3\sigma$ confidence level or more. The detectability of secondary lenses at different locations with respect to the source show strong dependence on the surface brightness and surface brightness gradient of the source. Comparing the blue and cyan regions in different subplots from left to right and top to bottom in each of the subgrids {\bf left} and {\bf right} also shows the strong dependence on the mass of the substructure as well as the observing frequency. Simulated observations of bands 8 and 9 are, in principle, of higher angular resolution compared to those of band 7. However, due to the high noise level, one can see that band 7 gives the best substructure detection prospects and band 9 provides the poorest conditions. The grid in subfigure {\bf left} shows the results with the phase errors included in the measurements of each frequency band. Subfigure {\bf right} shows the results without accounting for phase errors.}
\label{fig:pos_double_smooth}
\end{figure*}

\begin{figure*}
\includegraphics[width=0.45\textwidth]{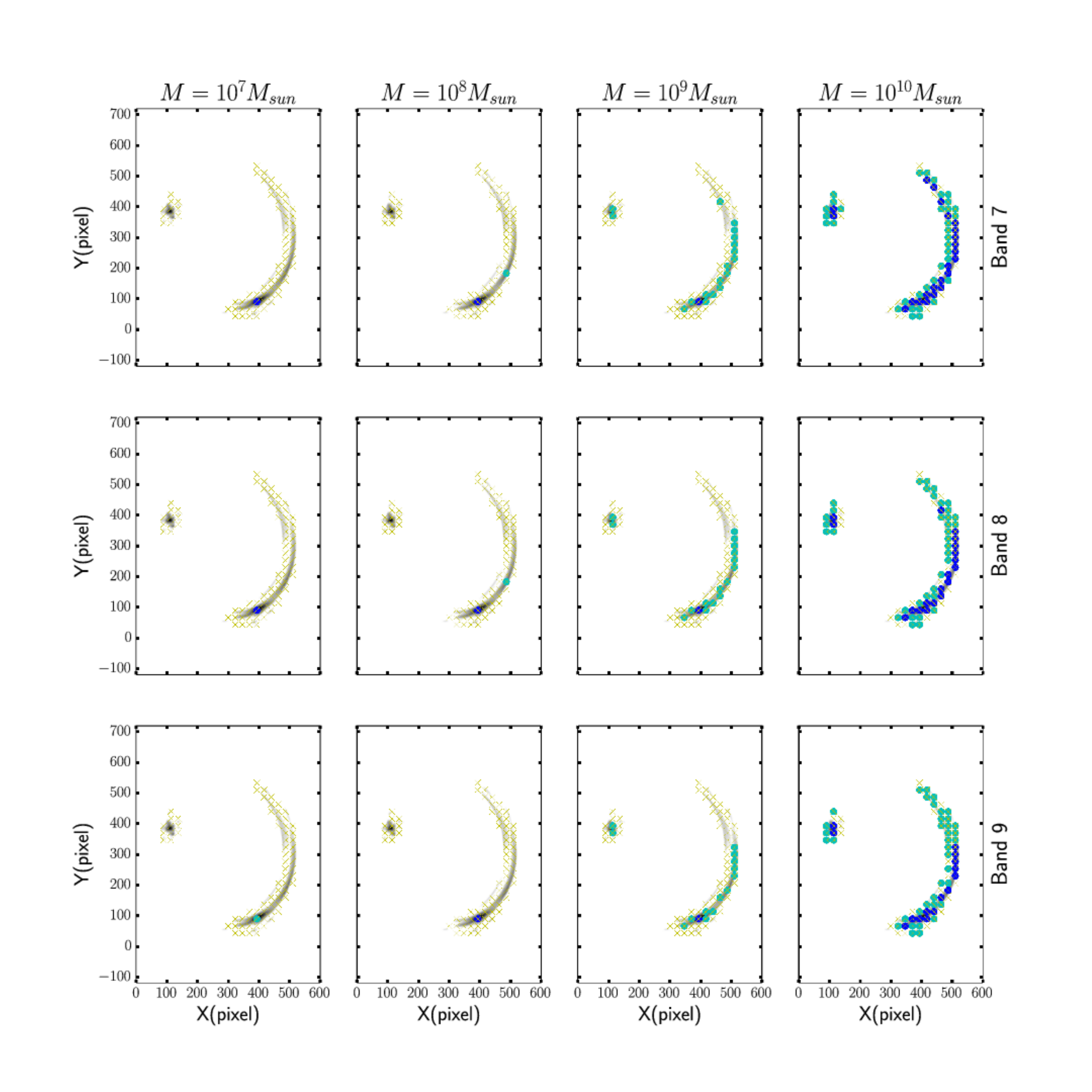}
\includegraphics[width=0.45\textwidth]{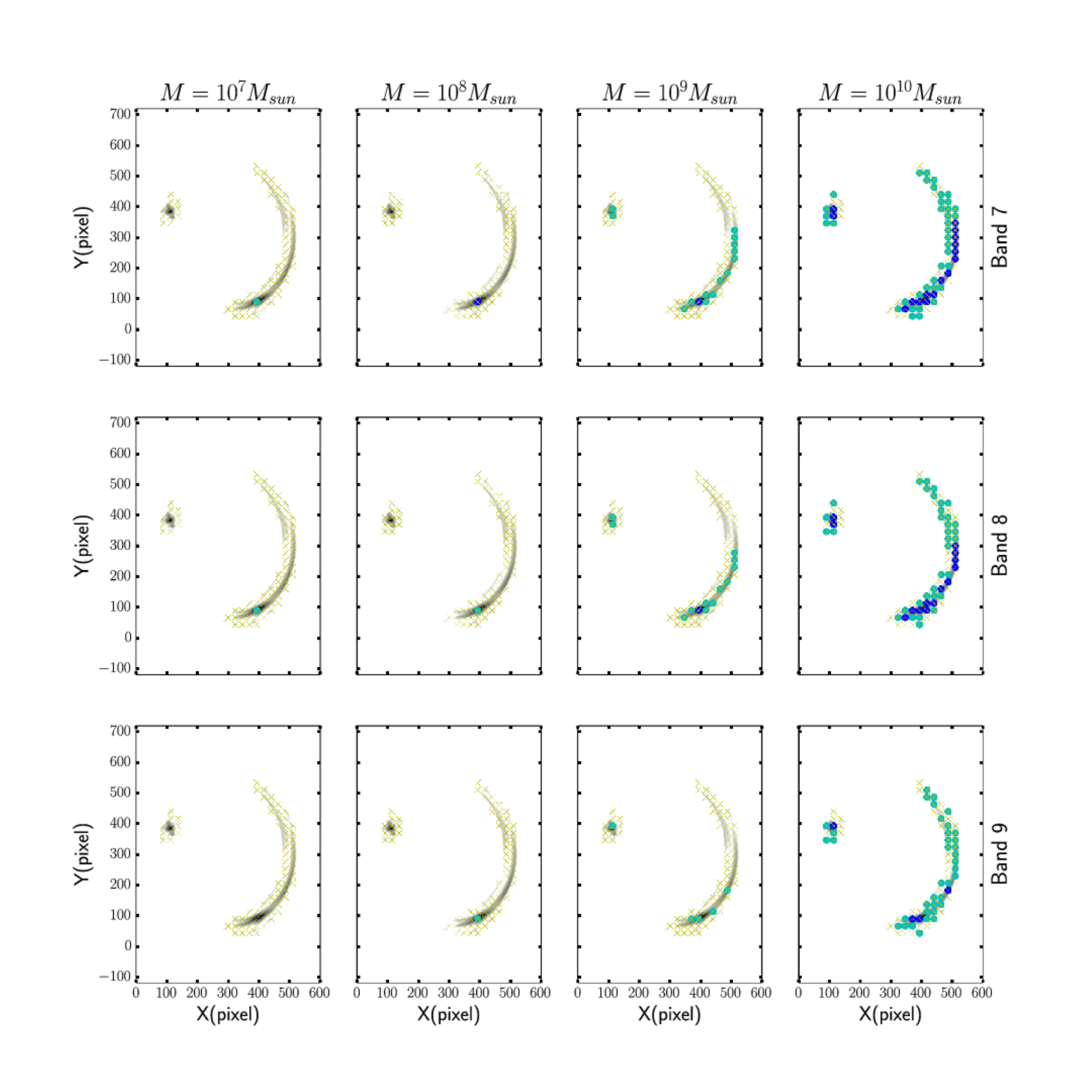}
\caption{The same as figure \ref{fig:pos_double_smooth}, but for the clumpy source model. Comparing the detection limits of the clumpy source for each combinations of subhalo mass--observing frequency confirms that even though possible in principle, detection of subhalos with masses lower than $\sim 10^9 M_\odot$ are challenging and strongly dependent on the image plane position of the lens with respect to the lensed images.}
\label{fig:pos_double_clumpy}
\end{figure*}

\begin{figure*}
\includegraphics[width=0.45\textwidth]{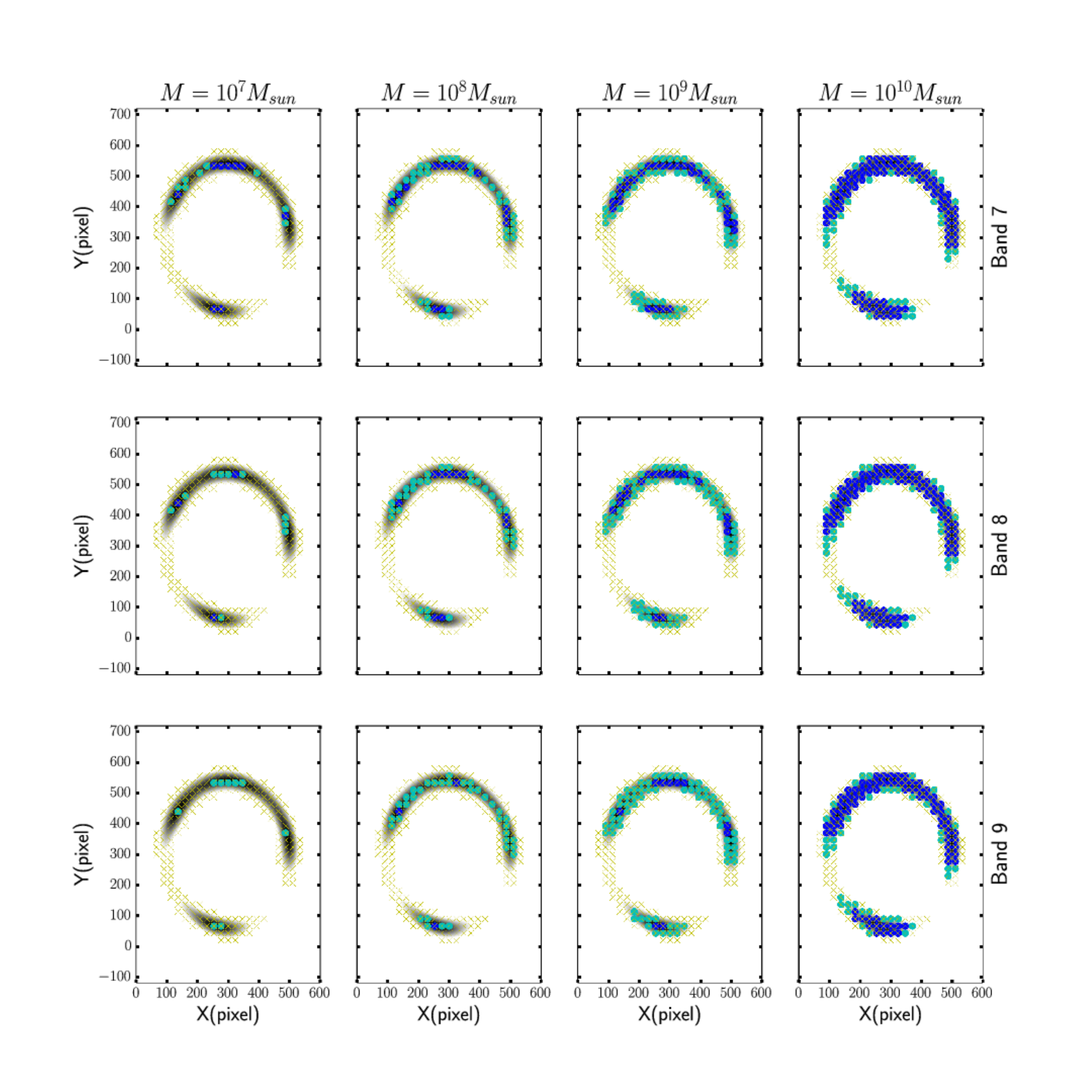}
\includegraphics[width=0.45\textwidth]{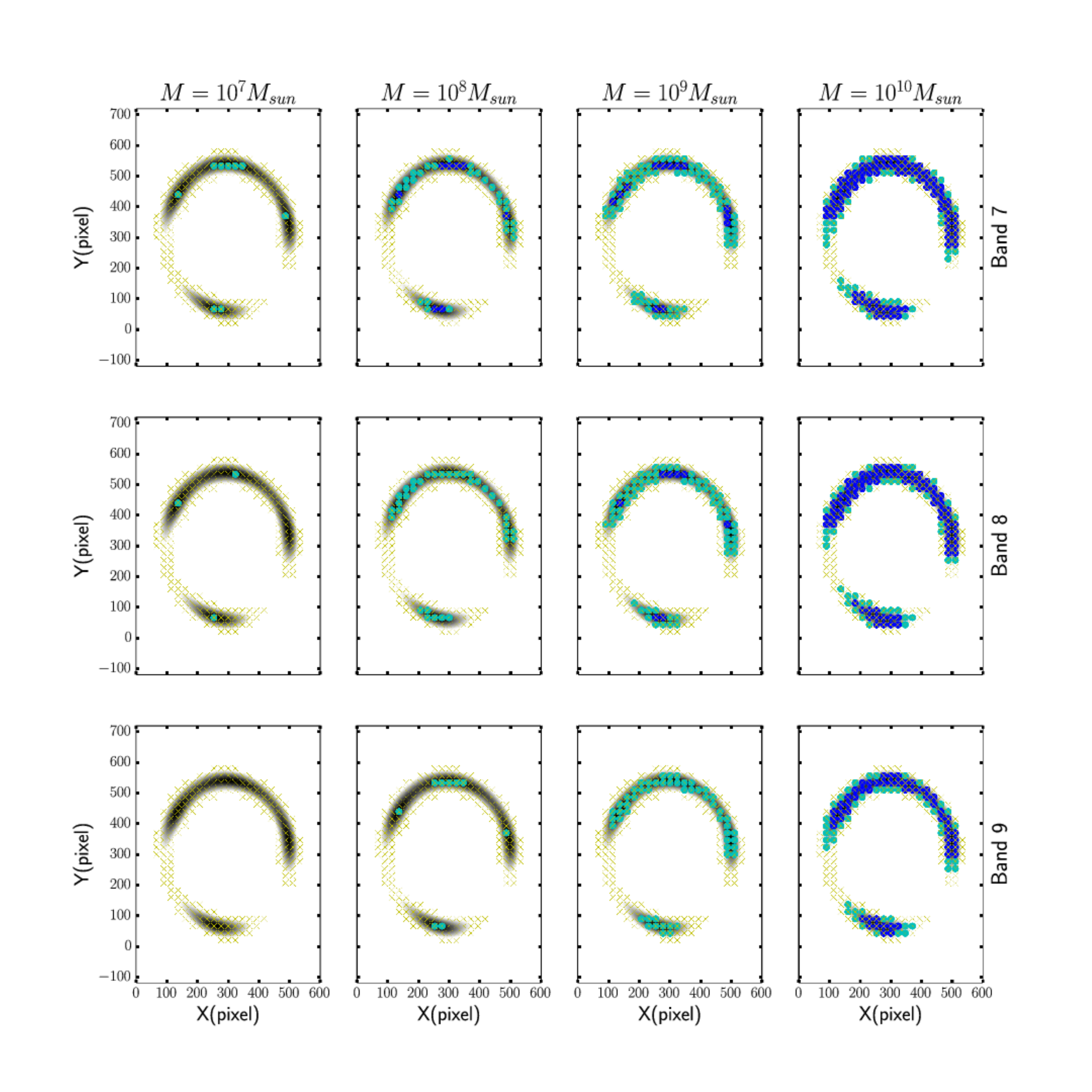}
\caption{The same as figure \ref{fig:pos_double_smooth}, for a fold lens configuration.}
\label{fig:pos_fold_smooth}
\end{figure*}

\begin{figure*}
\includegraphics[width=0.45\textwidth]{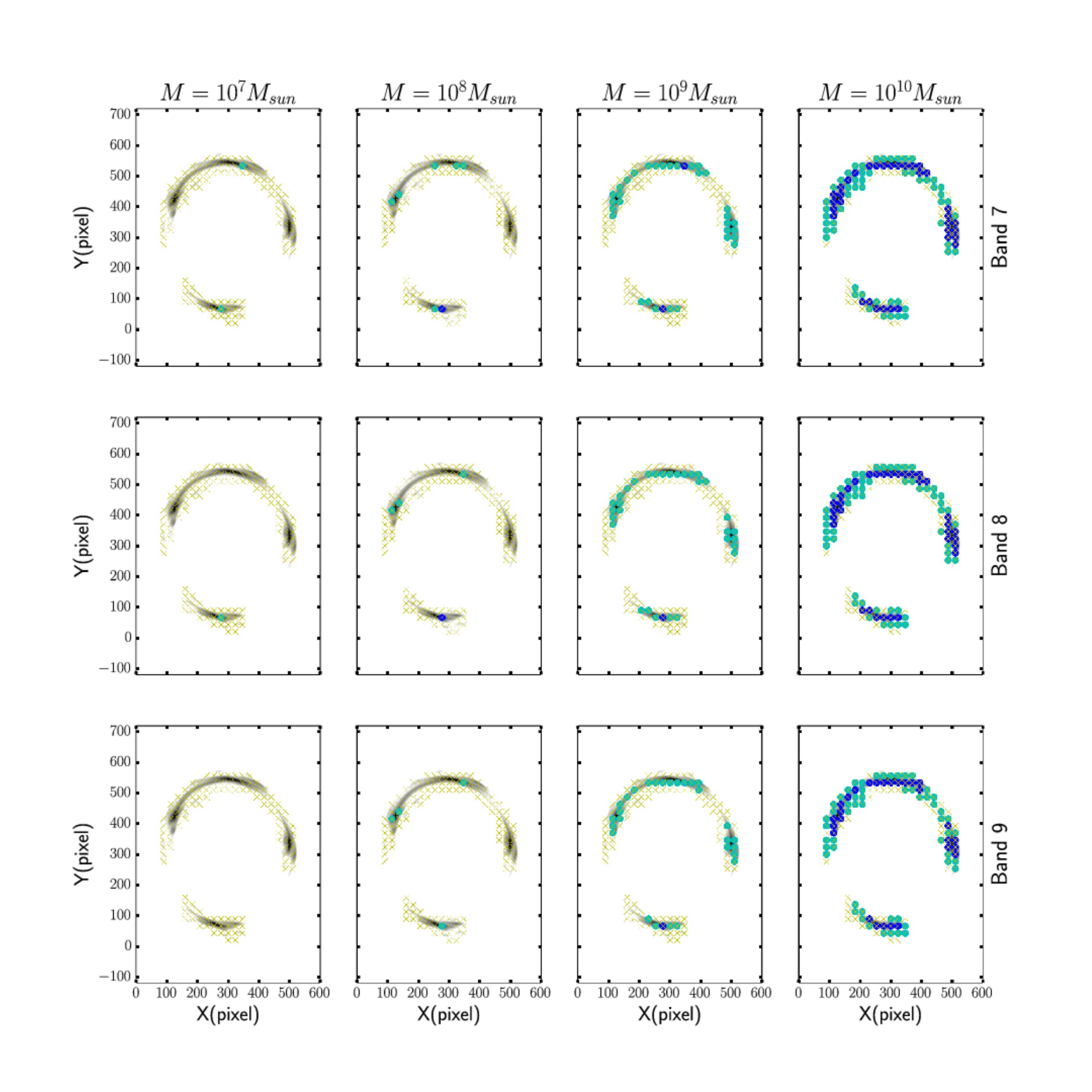}
\includegraphics[width=0.45\textwidth]{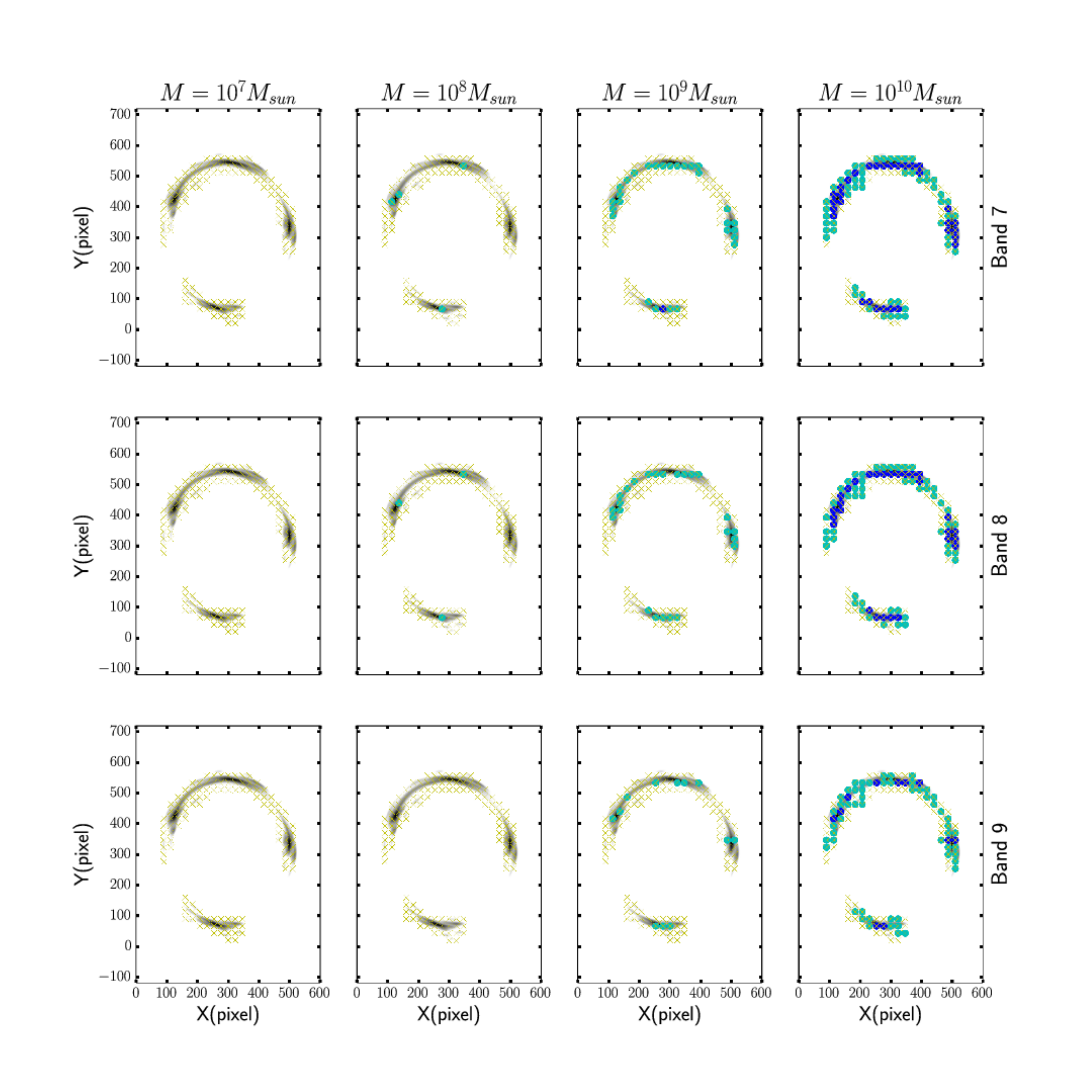}
\caption{The same as figure \ref{fig:pos_double_clumpy}, for a fold lens configuration.}
\label{fig:pos_fold_clumpy}
\end{figure*}

\begin{figure*}
\includegraphics[width=0.45\textwidth]{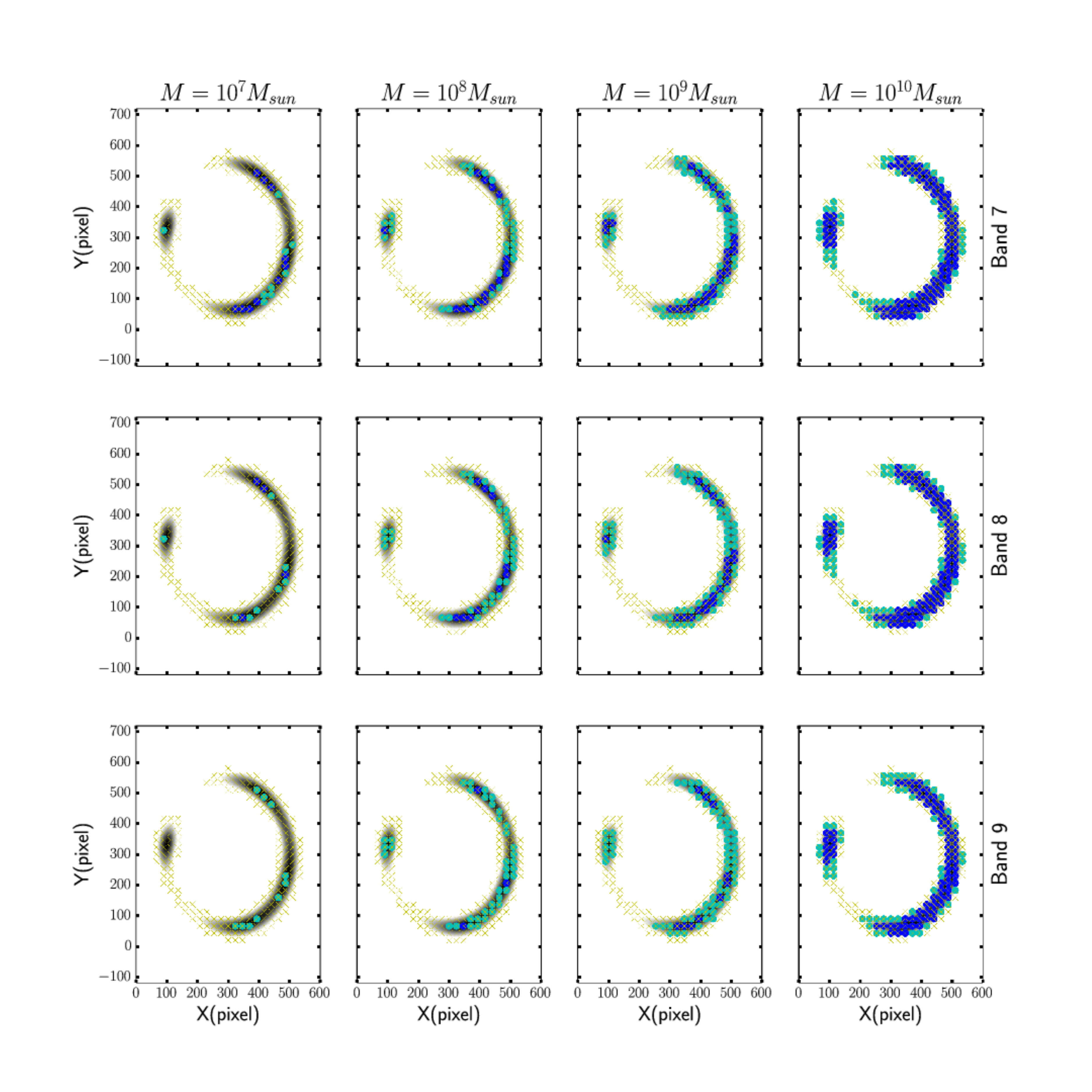}
\includegraphics[width=0.45\textwidth]{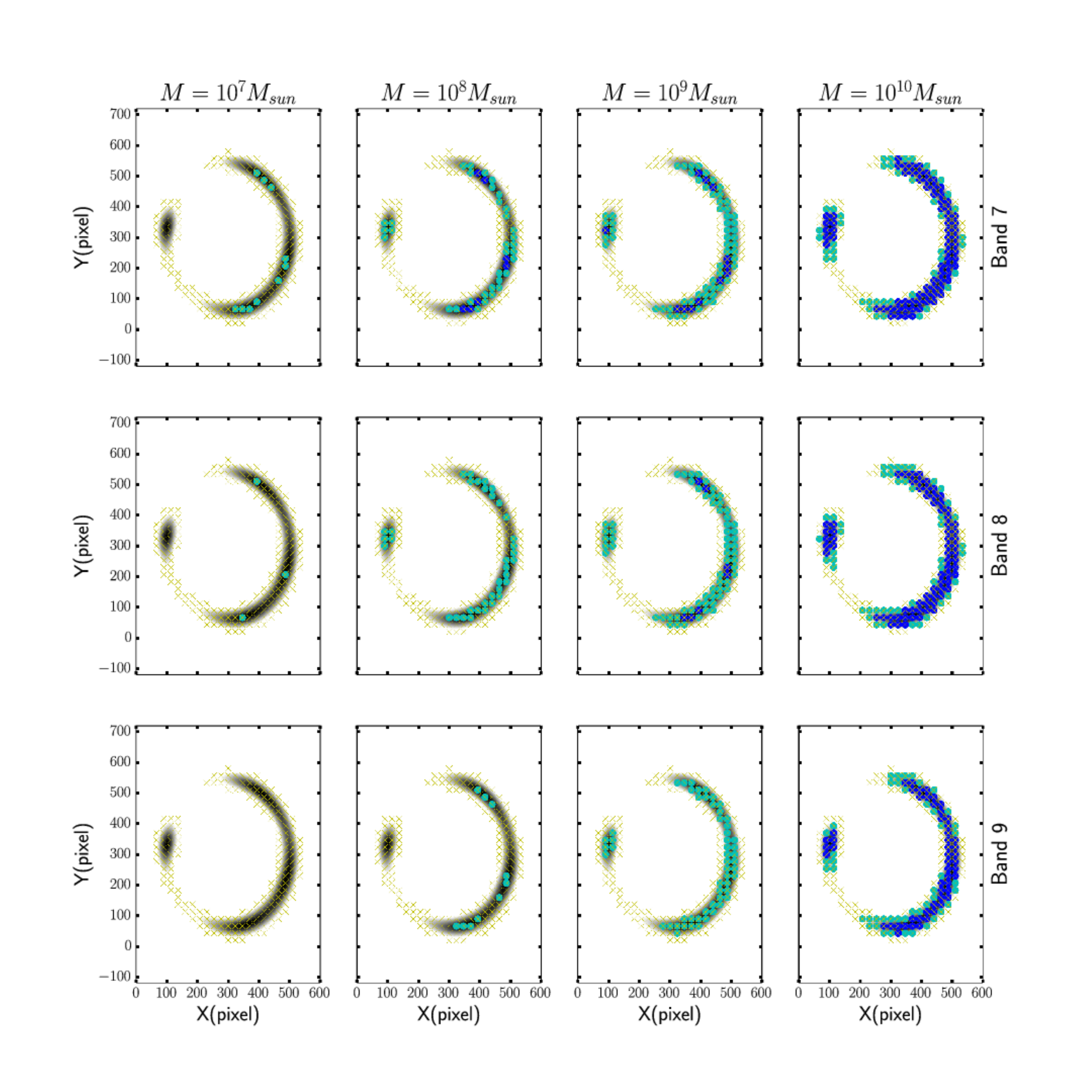}
\caption{The same as figures \ref{fig:pos_double_smooth} and \ref{fig:pos_fold_smooth}, for a cusp lens configuration.}
\label{fig:pos_cusp_smooth}
\end{figure*}

\begin{figure*}
\includegraphics[width=0.45\textwidth]{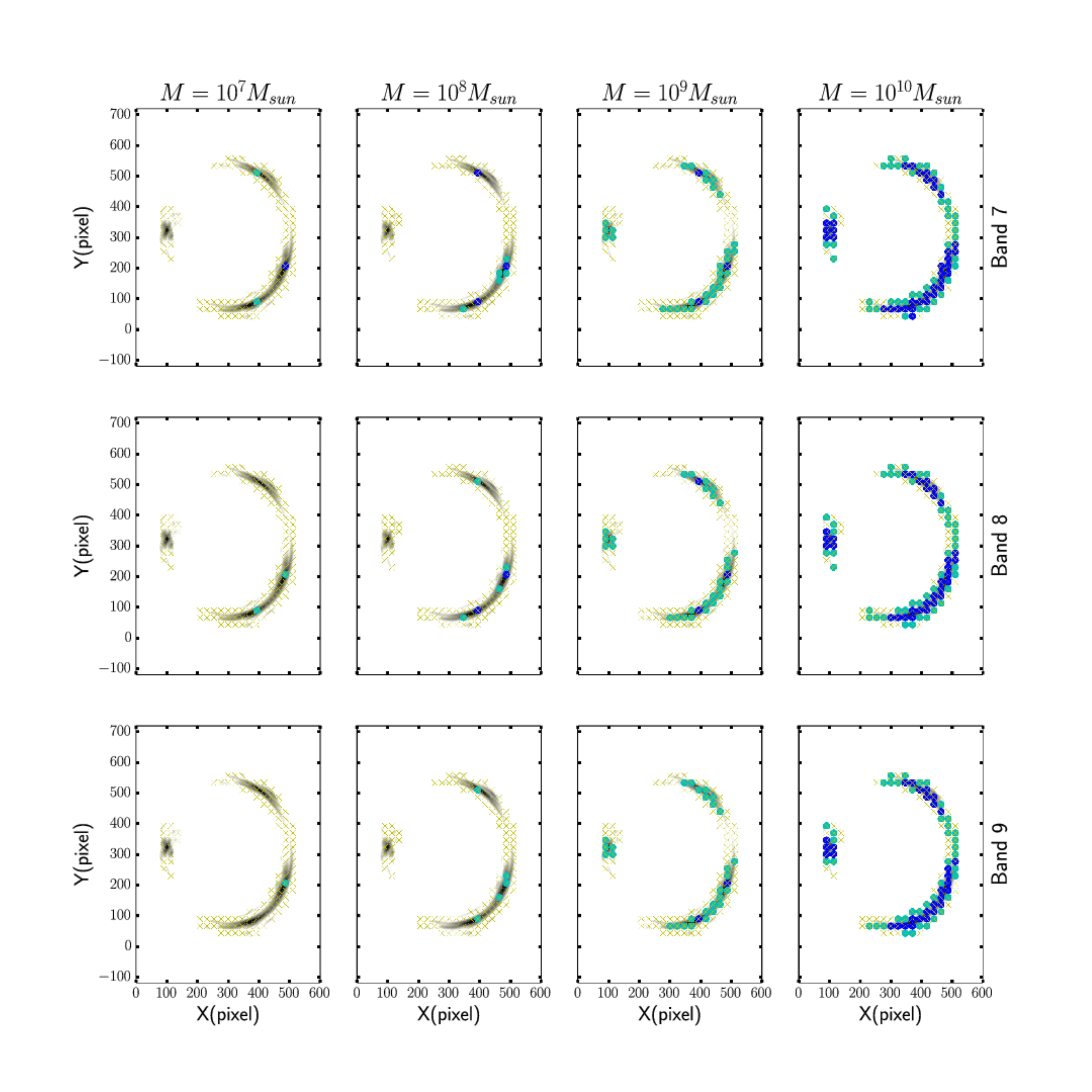}
\includegraphics[width=0.45\textwidth]{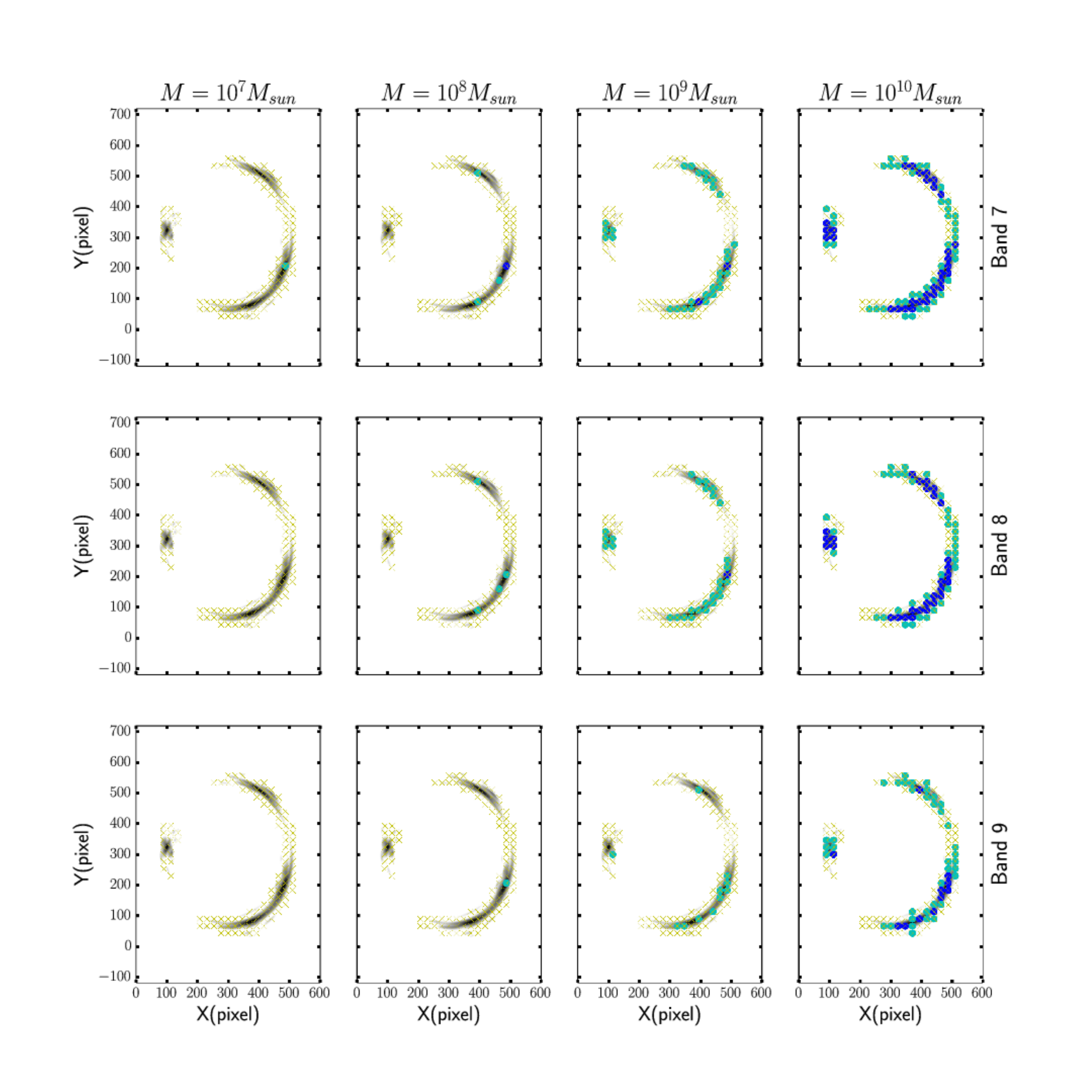}
\caption{The same as figures \ref{fig:pos_double_clumpy} and \ref{fig:pos_fold_clumpy}, for a cusp lens configuration.}
\label{fig:pos_cusp_clumpy}
\end{figure*}

In order to measure to what extent the presence, mass, and projected position of a lens perturber (subhalo) is recovered by our method, we assume a fixed density profile for the subhalo and investigate the detection probabilities at various positions as shown in Figure \ref{fig:sim_preparation}. The results of this process for the 6 different combinations of lens system configuration and source structure are shown in Figures \ref{fig:pos_double_smooth}, to \ref{fig:pos_cusp_clumpy}. All these figures follow the same pattern; Each row corresponds to a frequency band with bands 7, 8, and 9 from top to bottom. The confidence region for each mass step from $M_\mathrm{sub} = 10^7 M_\odot$ to  $M_\mathrm{sub} = 10^{10} M_\odot$ are shown in columns from left to right. The positions of yellow crosses in each subplot are where the test secondary lenses are placed. Dark blue regions are indicative of substructure positions at which a model with the correct mass, and position of the substructure if preferred over other models by $10 \sigma$ confidence level, can be detected while the regions in cyan show the positions at which the substructure detection is weaker ($\le 3\sigma$ confidence in recovered parameters).

A challenge to address is the degeneracy between the inner density slope and mass of the substructure, as previously discussed in the literature as well \citep[see e.g. ][]{VegettiandVog2014}. While all substructures of the same group (SIE/jaffe and NFW/Einasto) predict the position of the secondary lens correctly, the difference between predicted masses from different substructure models could be as large as an order of magnitude for the massive end ($M_\mathrm{sub} = 10^9 M_\odot$), even when parameters of the main lens and shear are fixed.

Cuspy, two--parameter density profiles such as SIS and pseudo-Jaffe are not well--supported by either observational fits to central density profiles of dwarf galaxies \citep[see e.g.][]{FloresandPrimack1994, Moore1994, Strigari+2006, Walker+2009, Boylan-kolchin+2011, Amorisco+2013, Collins+2014}, or simulated subhalos, which are, as discussed in section \ref{sec:different_forms_of_halo_substructure} best described by the Einasto profile with the shape parameter $\alpha_\mathrm{Ein}$ as the fitting parameter. However, to follow the common practice \citep[see e.g.][]{Metcalfandmadau2001, Metcalf2002, Vegetti+2010, Vegetti+2012, Hezaveh+2013b} in gravitational lens modeling to fit SIS or pseudo--Jaffe to the lens substructure, we report here the mass detection limit given by our method for these density profiles. Figures \ref{fig:pos_double_smooth} to \ref{fig:pos_cusp_clumpy} illustrate the detection significance of pseudo--Jaffe perturbers of various $M_\mathrm{vir}$ as a function of the projected position of the substructure with respect to the lens images.

With all the parameters of the main lens, the substructure, and the source left to fit simultaneously, neither the mass, nor position of the secondary lens can be well--recovered, even though models with substructure are preferred to those without. We employ the following strategy to help the optimization algorithm to avoid local minima. First, we find the best solution to the system when all parameters (including source parameters) are considered free. This step is given at least 10000 iterations to converge on the mass of the main lens. In the next step, we use these best fit parameters of the main lens together with their uncertainties as priors to find the mass and position of the substructure. The first step helps to constrain the parameters related to the main lens such as the mass and position of the lens, and the position, size and integrated flux of the source. Consequently, the search space is reduced for less influential parameters such as the mass and position of the secondary lens as well as the external shear inside the main lens.

The illustrated detections are derived from modeling simulated ALMA observations with a substructure of a specific mass placed in the vicinity of lens images as detailed in section \ref{sec:method}. The dark blue points in each figure show the positions where the mass and position of the substructure are recovered at $10\sigma$ confidence through the process explained in section \ref{sec:method}. The cyan regions correspond to the subhalo positions at which the mass and position of the secondary lens is correctly recovered with $3\sigma$ confidence level or better. In the set of simulations using \emph{smooth} source model, the source and the inner density profile of the secondary lens are fixed, while all other parameters corresponding to the primary and secondary lens are free. The spatial distribution of blue and cyan grids show strong dependence on the surface brightness and surface brightness gradient of the source, as expected. This is consistent with our results using a \emph{smooth} source compared to the corresponding results using the \emph{clumpy} source model, where the most--favorable regions for the secondary lens is limited to the brightest regions of the image plane which are smaller than those of the \emph{smooth} source simulations by $\sim 40$ percent. 

Comparing the blue and cyan regions in different subfigures from left to right and top to bottom also shows strong dependence on the mass of the substructure as well as the observing frequency. Simulated observations of bands 8 and 9 are in principle of higher angular resolution compared to those of band 7. However, due to the high noise level (due to phase instability), one can see that band 7 gives the best substructure detection prospects, and band 9 provides the poorest conditions. These observations are also consistent with what is previously presented in the literature \citep[see e.g. ][]{Kochanek2004, SchneiderandSluse2013, Vegetti+2014}. This is illustrated in comparison between the left and right grids in all figures \ref{fig:pos_double_smooth} through \ref{fig:pos_cusp_clumpy} where the left side grid are made by correcting for the measurement error in phases of the simulated visibilities.  The right side grid is included to show the extent to which phase errors in each frequency band can affect the recovery of subhalo parameters.

%% =========================== %%
%%          DISCUSSION         %%
%% =========================== %%
\section{Discussion}
\label{sec:discussion}

\subsection{Substructure mass estimation}
\label{subsec:substructure_mass_estimation}
In this section we argue that comparison between the mass within the Einstein radius of the lens, derived directly from solving the lensing equation, and the physical mass of the lens, i.e. the tidal subhalo mass as discussed in N--body simulations, strongly depends on the assumptions about both the density profile and truncation radius of the halo.

Conventionally, simulated subhalo mass is presented as $M_\mathrm{tidal}$, while the dwarf galaxy total/integrated mass is described within the central 300 pc as $M_{300}$. When discussing the detection limit of dark subhalos via gravitational lensing in the context of galaxy formation, it is crucial to make sure that the subhalo mass definitions assumed in lens modelings and N--body simulations are properly converted. This issue can be reduced to two different interpretations of the truncation radius of the halo, which becomes especially important when it comes to low--mass halos where detailed lensing signals are strongly affected by inner density slope and concentration \citep[][]{Zackrisson+2008}. 

Einstein radius is defined as the radius within which, the mean surface mass density of the lens is equal to the critical surface mass density of the Universe at the redshift of the lens. Consequently, the Einstein mass is defined as the mass within the Einstein radius of the lens,
\begin{equation}
M_\mathrm{Ein} = \Sigma_\mathrm{crit} \uppi R_\mathrm{Ein}^2 \nonumber
\end{equation}

In a strong lens system, the Einstein radius is a direct measure of the Einstein ring, arc or image separation. However, in lens systems such as the compound systems discussed in this paper, the direct measurement of the size of the Einstein radius using image separation is limited by the angular resolution. Additionally, the projected mass contributing to the surface mass density of the lens corresponds to a combination of mass components including the main lens, the lens perturber (substructure), and an external shear accounting for the remaining mass components on the lens plane or along the line of sight.

\begin{figure}
\includegraphics[width=0.48\textwidth]{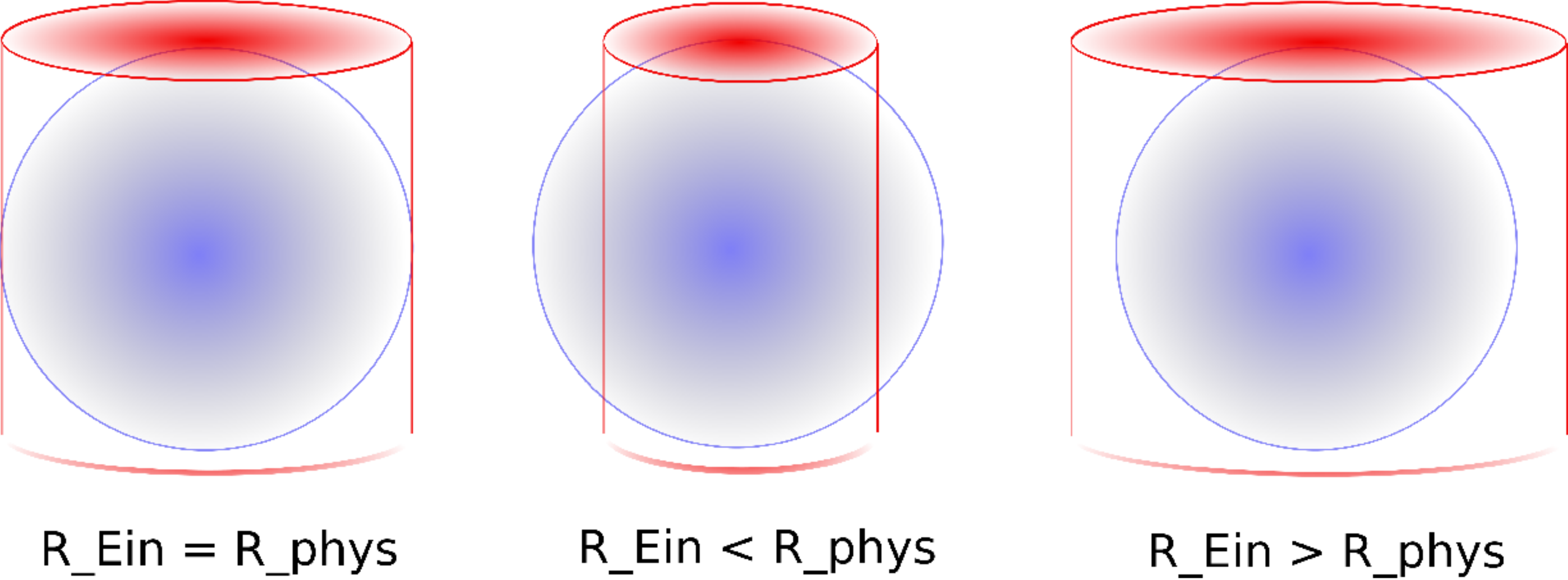}
\caption{This simplified illustration shows how deriving the 3D mass of a spherical halo based on the mass measured in gravitational lensing (projected mass) could vary depending on the assumed physical radius of the halo.} 
\label{fig:mass_illustration}
\end{figure}

\begin{figure*}
\includegraphics[width=0.9\textwidth]{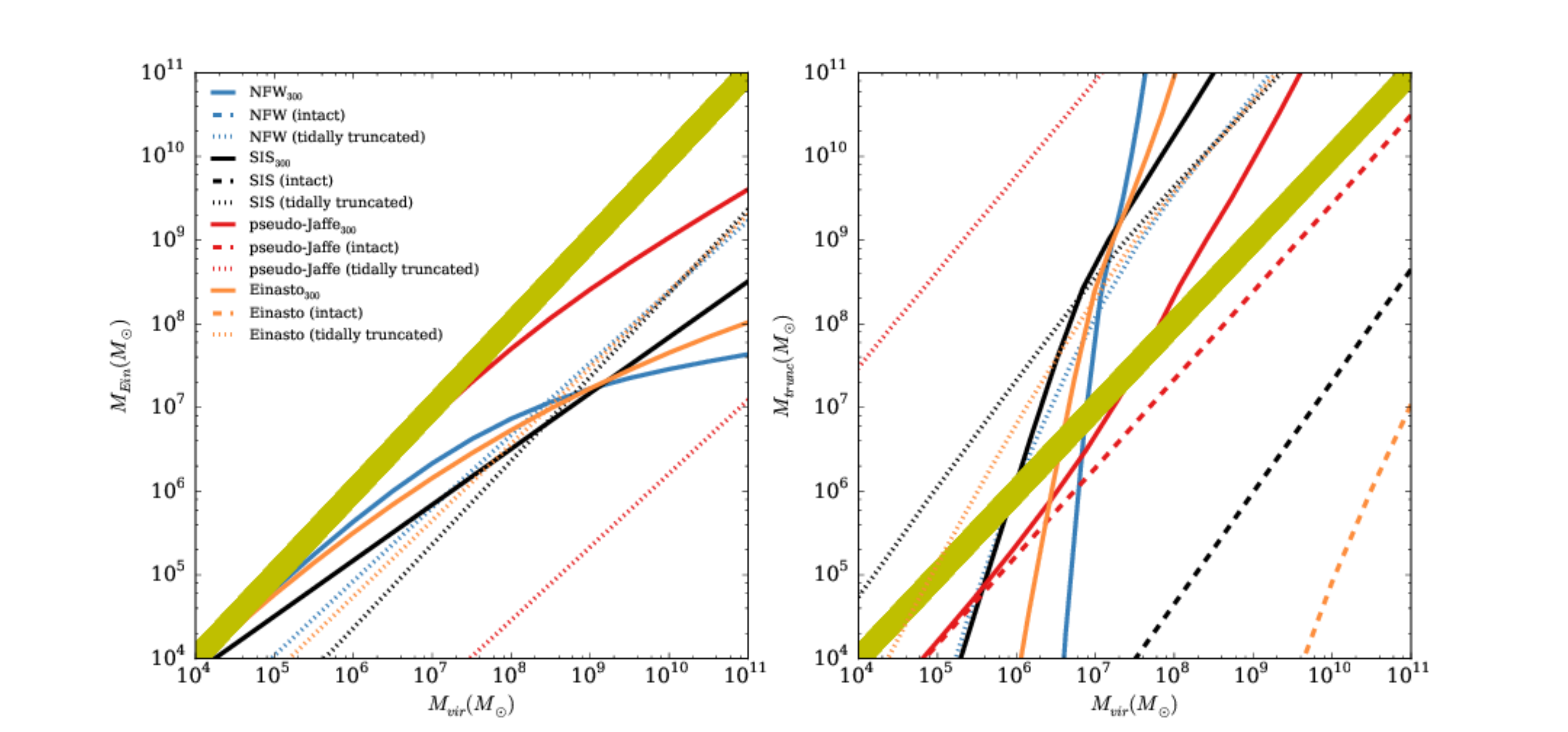}
\caption{The importance of consistent mass conversion in comparing masses derived from lens models and N--body simulations and the effect of accounting for tidal stripping effects in outer regions of subhalos as opposed to field halos. {\bf Left:} Translation between the virial mass and the truncated mass of halos at $z=0.5$, for different inner density profiles. The thick yellow line has a slope of one and is therefore where all intact halos overlap. {\bf Right:} Translation between the truncated mass and the mass within the Einstein radius of halos at $z=0.5$, assuming $R_\mathrm{trunc} = R_\mathrm{Ein}$. Deviation of the mass within the half--light radius ($r_\mathrm{trunc}=300$ pc) from the virial mass is clearly visible for halos with $M_\mathrm{vir}>\sim 10^6 M_\odot$ with NFW and Einasto inner density profiles while the difference tends to grow monotonously for halos with steeper logarithmic slopes, i.e. SIS and pseudo--Jaffe. One needs to keep in mind that halos with $M_\mathrm{vir}\gtrsim 10^9 M_\odot$ can no longer be considered in the context of galactic substructure and consequently dwarf galaxies. The right panel shows how the mass measured within the Einstein radius deviates from the actual halo mass within the truncation radius (the virial radius for intact halos). The first immediate result is the group of dotted lines (indicating intact halos) in the lower right corner of the plot, implying large deviation of the deprojected mass within the Einstein radius from the subhalo mass within the virial radius. The red line, corresponding to pseudo--Jaffe profile shows the least deviation, while the NFW inner profile gives rise to a total Einstein mass of smaller than $10^4 M_\odot$ for subhalos with $M_\mathrm{vir}\leq 10^{11} M_\odot$ consistent with expectations by comparing the central density slopes in figure \ref{fig:profiles}.} 
\label{fig:mass_comparison}
\end{figure*}

Figure \ref{fig:mass_illustration} illustrates how the measured $M_\mathrm{Ein}$, depends on the mass concentration within the lens as well as its physical truncation radius. In other words, the degeneracy between the Einstein mass of an observed lens and its physical 3D mass is only breakable if all fitting parameters of the model halo ($M_\mathrm{trunc}$, $R_\mathrm{trunc}$, $M_\mathrm{Ein}$ and $R_\mathrm{Ein}$) are known. The illustration in figure \ref{fig:mass_comparison} shows that when $R_\mathrm{Ein}<R_\mathrm{phys}$ (the case for virial halo of all the covered mass range), $M_\mathrm{Ein}$ derived from gravitational lensing only contains part of the physical mass content of the lens. Therefore, in order to derive the 3D mass content of the lens, one needs to assume a 3D density profile as well as a truncation radius within which the actual mass content of the lens is meaningful.

While in cases where $R_\mathrm{Ein}<R_\mathrm{trunc}$ (the case for $M_\mathrm{300}$), the mass distribution beyond $R_\mathrm{Ein}$ only affects the macrosolution, i.e. the positions of multiple images which are beyond the resolution limit for low--mass subhalos.

Therefore, in order to derive the 3D mass of the lens the truncation radius $R_\mathrm{trunc}$, total mass within $R_\mathrm{trunc}$, and the mass distribution within the lens are to be fitted simultaneously. 

The results presented in section \ref{sec:results} deal with three dimensional substructure 3D tidal masses $\sim 10^7 - 10^{10} M_\odot$. The corresponding mass enclosed within the Einstein radius of each halo can be seen in figure \ref{fig:mass_comparison} indicating that using the full ALMA array with a 10 km maximum baseline (about 10 times better spatial resolution than discussed in \citet{Hezaveh+2013b}) enhances the mass sensitivity by more than an order of magnitude.

\subsection{Dark matter substructure as lens perturbers}
\label{subsec:current_state_of_dark_matter_substructure_as_lens_perturbers}
The state of detecting dark substructure in strong lenses depends on the alignment of the system as well as the angular resolution of the observations. The trade--off between the source size and observational resolution determines the instruments available for hunting dark subhalos.

\citet {Vegetti+2010, Vegetti+2012}, using the Hubble space telescope and the Keck telescope, report the detection of two dark substructures of $M_\mathrm{Ein} = 3.5\times10^9 M_\odot$ and $M_\mathrm{Ein} = 1.9\times10^8 M_\odot$ in two strong lens systems SDSS J0946+1006 and JVAS B1938+666. They derive the gravitational potential of the lens and the \emph{local} residual of the potential of the perturbed system compared to that of a smooth lens model, to constrain the projected mass of the perturber. According to the authors, the masses of substructures detected in these systems correspond to $M_{300}^\mathrm{p-Jaffe} = 3.0 \times 10^8 M_\odot$ and $M_{300}^\mathrm{SIS} = 3.3 \times 10^7 M_\odot$ for the more massive substructure and $M_{300}^\mathrm{p-Jaffe} = 1.1 \times 10^8 M_\odot$ and $M_{300}^\mathrm{SIS} = 6.1 \times 10^6 M_\odot$ for the less massive substructure. However, as discussed in \ref{subsec:substructure_mass_estimation}, in order to reproduce the deprojection one requires the physical truncation mass assumed for the deprojection as well as the Einstein mass and radius.

More recently \citet[][]{Birrer+15} presented a more generic method, using Monte Carlo algorithms, to solve simultaneously for substructures in the lens and source plane. They reconstruct the surface brightness by finding the global minimum of the $\chi^2$ and achieve a lens substructure resolution of $\sim 10^8 M_\odot$ within the HST data.

A similar work to the one presented in this paper is published by \citet{Hezaveh+2013b} to detect dark subhalos in compound lens systems using ALMA. They use spatially resolved spectroscopy of high excitation (CO) lines tracing the cores of star--forming regions in lens systems of different alignments.  In their simulations a dusty star forming galaxy at $z=2$ is lensed by a foreground galaxy halo at $z=0.5$ containing dark halo substructures. They use a total of 1 hr integration time with band 7 ALMA cycle 1 configuration (maximum baseline = 1.1 km). The detection mass limit for single subhalos in their simulations is reported to be $M_\mathrm{Ein} > 10^8 M_\odot$, translating to $M_\mathrm{vir}^\mathrm{SIS} = 3\times 10^{10} M_\odot$ for a pseudo--Jaffe subhalo with, $\rho(r) \propto r^{-2}$ inner profile (Figure \ref{fig:mass_comparison}). It is worth mentioning that the standard dark matter density profiles are not considered in any of the above work making it difficult to derive the link between observations and theory.

\subsection{Low redshift line--of--sight contamination and external shear vs. central density profile}
\label{subsec:low_redshift_line_of_sight_contamination_and_external_shear_vs_central_density_profile}
One often--neglected lensing effect in galaxy--galaxy lensing is that the lensed image is subject to more than a single lens at a single redshift. The probability of multi--plane lensing strongly depends, among other things, on the redshift of the source. This is often accounted for --implicitly-- by adding external shear and convergence as free parameters of the individual lens model. With improving methods and tools in lens measurements, the effect of line--of--sight (LOS) lensing may become a dominant factor that needs precise treatment. There are many papers discussing these methods \citep[see e.g. ][]{Keeton+1997, Keeton2003, Keeton+2004, Gavazzi+2012, Schneider2014, Lee+2014, McCully+2016}. \citet[][]{Wambsganss+2004} find that in 95$\%$ of the cluster lens systems with $z_s=1$, the secondary lens effect is negligible. While this fraction decreases steeply to 68$\%$ for systems with $z_s=3.8$. Whether the line--of--sight contamination effects occur within the strong or weak lensing regime, precise modeling of a multi--plane lens system is impossible, because of the large number of unknown parameters. A statistical approach is, however, possible to take within the context of the background cosmology. One aspect of this correction is assuming a cosmic shear effect in addition to the simple lens model. \citet{Schneider2014} discusses the significance of lens perturbers in intermediate redshifts on cluster lens systems, in contrast to galaxy--scale lens systems where the alignment of multiple lens planes at different redshifts is improbable, although not impossible \citep[e.g.][]{Chae+2001, Gavazzi+2008}. In the strong lensing regime, accounting for the main lens gives a good approximation of the lens model and other massive objects along, or close to, the line of sight add a lensing effect that can be as strong as a few percent of the strong lens. However, galaxy--galaxy lens systems with more than one deflector along the line of sight are less probable than a cluster lens system of the same redshift combination.

Therefore, results of the small--scale strong lensing effects discussed in this paper are not likely to be due to line--of--sight contamination. In an unlikely case where the substructure lensing is due to a secondary lens at a redshift different than the primary lens, this gives an estimate of the systematic error on the subhalo mass fraction and mass function derived from a statistically--meaningful sample of such detections.

%% =========================== %%
%%          SUMMARY            %%
%% =========================== %%
\section{Summary}
\label{sec:summary}
We use simulations of strong galaxy--galaxy lensing systems as seen with ALMA in band 7, 8, and 9 in search of completely dark or high mass--to--light ratio subhalos through small--scale flux density perturbations in a single lensed image. All simulations use the \emph{ALMA Full ops most extended} configuration. We aim for lens perturbers of $M_\mathrm{sub} = 10^7 - 10^{10} M_\odot$ in simulated galactic dark matter halos at $z\sim0.5$. The lensed source, i.e. the sub--mm galaxy, is at $z\sim2.0$, a typical redshift for lensed SMGs. The analysis is done on simulated complex visibilities that are weighted differently to put emphasis on different angular scales of the flux density of the source. 

We show that given a single perturber within the lens system with macroimage separation of $\sim 1$ arcsec and $\mu \approx 10$, and the presence of at least one more magnified image in the system, pseudo--Jaffe perturbers more massive than $M_\mathrm{vir} = 10^7 M_\odot$ will be detectable with 2hr observations using ALMA band 7, band 8, and band 9. We also show that the detection limit strongly depends on the projected position of the subhalo in the image plane, with the brightest regions being the most promising. However, the detection significance could still be high for the fainter regions of the lens image, depending on the observing frequency and mass of the substructure. Moreover, we argue that while the recovered position of the lens perturber is robust, the estimated mass highly depends on the details of the inner mass profile used for model selection.

There are other parameter choices that affect the detection limit, both in mass and position of the secondary lens.  We discuss the effect of internal source structure -- from smooth single--component galaxies to clumpy multi--component galaxies with structures on the scale of secondary lens image separation. We also investigate the effect of accounting for phase errors, which carry the sensitive information about the correct position of the secondary lens.
\section*{Acknowledgements}
E.Z. acknowledges funding from the Swedish Research Council (project number 2011--5349) and the Wenner--Gren foundations.

%%%%%%%%%%%%%%%%%%%%%%%%%%%%%%%%%%%%%%%%%%%%%%%%%%

%%%%%%%%%%%%%%%%%%%% REFERENCES %%%%%%%%%%%%%%%%%%

% The best way to enter references is to use BibTeX:

%\{mnras}
%\bibliography{example} % if your bibtex file is called example.bib

% Alternatively you could enter them by hand, like this:
% This method is tedious and prone to error if you have lots of references

%%%%%%%%%%%%%%%%%%%%%%%%%%%%%%%%%%%%%%%%%%%%%%%%%%

% Don't change these lines
\bsp	% typesetting comment
\label{lastpage}
\end{document}